\documentclass[12pt]{iopart}
\usepackage{iopams}  
\usepackage{bm}
\begin{document}
\newcommand{\s}{{\sigma}}
\newcommand{\rv}{{\bf r}}
\newcommand{\xv}{{\bf x}}
\newcommand{\kv}{{\bf k}}
\newcommand{\eps}{{\varepsilon}}
\newcommand{\uv}{{\bf u}}
\newcommand{\vv}{{\bf v}}
\newcommand{\qv}{{\bf Q}}
\newcommand{\Qv}{{\bf Q}}
\newcommand{\jv}{{\bf j}}
\newcommand{\Sv}{\vec{S}}
\newcommand{\Mv}{\vec{M}}
\newcommand{\nv}{\vec{n}}
\newcommand{\mv}{\vec{m}}
\newcommand{\nh}{{\hat n}}
\newcommand{\qh}{{\hat Q}}
\newcommand{\mh}{{\hat m}}
\newcommand{\lh}{{\hat \ell}}
\newcommand{\Sh}{{\hat S}}
\newcommand{\oDelta}{{\overline{\Delta}}}
\def\mm#1{\underline{\underline{{#1}}}}
\newcommand{\um}{{\mm{u} }}
\newcommand{\Lm}{{\mm{\Lambda} }}
\newcommand{\pv}{{\bf p}}
\newcommand{\px}{{\partial_x}}
\newcommand{\py}{{\partial_y}}
\newcommand{\ppi}{{\partial_i}}
\newcommand{\ppj}{{\partial_j}}
\newcommand{\pz}{{\partial_z}}
\newcommand{\ppk}{{\partial_k}}
\newcommand{\ppl}{{\partial_l}}
\newcommand{\oh}{{\frac{1}{2}}}
\newcommand{\nsm}{{n$\sigma$m\ }}
\newcommand{\tRe}{{\text{Re}}}
\newcommand{\cH}{{\mathcal H}}
\newcommand{\cL}{{\mathcal L}}
\def\rf#1{(\ref{#1})}
\def\rfs#1{Eq.~\rf{#1}}
\newcommand{\BdG}{Bogoliubov--de~Gennes}
\newcommand{\grad}{{\bm{\nabla}}}
\newcommand{\zh}{\hat{z}}
\newcommand{\xh}{\hat{x}}
\newcommand{\Uh}{\hat{U}}
\newcommand{\yh}{\hat{y}}
\newcommand{\bh}{\hat{b}}
\newcommand{\ch}{\hat{c}}
\newcommand{\alphah}{\hat{\alpha}}
\newcommand{\Qh}{\hat{Q}}
\newcommand{\Psih}{\hat{\Psi}}
\newcommand{\Qb}{\bar{Q}}
\newcommand{\Deltab}{\bar{\Delta}}
\newcommand{\bx}{{\bf x}}
\newcommand{\bR}{{\bf R}}
\newcommand{\be}{\begin{equation}}
\newcommand{\ee}{\end{equation}}
\newcommand{\bea}{\begin{eqnarray}}
\newcommand{\eea}{\end{eqnarray}}
\newcommand{\bse}{\begin{subequations}}
\newcommand{\ese}{\end{subequations}}
\newcommand{\oct}{\Omega_{{\rm c}2}}
\newcommand{\bv}{\bar{{\bf v}}}
\newcommand{\bu}{{\bf u}}
\newcommand{\rG}{{\rm G}}
\newcommand{\rM}{{\rm M}}
\newcommand{\chip}{\chi_{\rm P}}
\newcommand{\phaseshift}{\delta_{s}}
\newcommand{\deltabcs}{\Delta_{\rm BCS}}
\newcommand{\deltahbcs}{\Deltah_{\rm BCS}}
\newcommand{\Hcz}{H_c^{\rm Z}}
\newcommand{\mubohr}{\mu_{\rm B}}
\newcommand{\psil}{\psi^{<}}
\newcommand{\psig}{\psi^{>}}
\newcommand{\phil}{\phi^{<}}
\newcommand{\phig}{\phi^{>}}
\newcommand{\sech}{{\rm sech}}
\newcommand{\rmc}{{\rm c}}
\newcommand{\nf}{n_{\rm F}}
\newcommand{\as}{a_{\rm s}}
\newcommand{\thetad}{\Theta_{\rm D}}
\newcommand{\td}{\tilde{\Delta}}
\newcommand{\te}{\tilde{\epsilon}_{\rm F}}
\newcommand{\ec}{\varepsilon_{\rm c}}
\newcommand{\fermiint}{\lambda}
\newcommand{\Ec}{E_{\rm c}}
\newcommand{\vf}{v_{\rm F}}
\newcommand{\vd}{v_{\Delta}}
\newcommand{\je}{{\bf j}_{\rm e}}
\newcommand{\ksl}{\not\!k}
\newcommand{\asl}{\not\!a}
\newcommand{\psl}{\not\!p}
\newcommand{\qsl}{\not\!q}
\newcommand{\delsl}{\not\!\partial}
\newcommand{\omd}{\omega_{\delta}}
\newcommand{\curD}{{\cal D}}
\newcommand{\curA}{{\cal A}}
\newcommand{\curH}{{\cal H}}
\newcommand{\vkk}{V_{{\bf k},{\bf k}'}}
\newcommand{\tk}{T_{\rm K}}
\newcommand{\jk}{J_{\rm K}}
\newcommand{\jknought}{J_{{\rm K}0}}
\newcommand{\tc}{{T_{\rm c}}}
\newcommand{\tco}{{T_{\rm c0}}}
\newcommand{\xf}{\xi_{\phase}}
\newcommand{\phdag}{{\phantom{\dagger}}}
\newcommand{\ps}{{\phantom{*}}}
\newcommand{\vol}{V}
\newcommand{\pf}{p_{\rm F}}
\newcommand{\tkf}{\tilde{k}_{\rm F}}
\newcommand{\ef}{\epsilon_{\rm F}}
\newcommand{\tf}{T_{\rm F}}
\newcommand{\kf}{k_{\rm F}}
\newcommand{\kfup}{k_{{\rm F}\uparrow}}
\newcommand{\kfdown}{k_{{\rm F}\downarrow}}
\newcommand{\curG}{{\cal G}}
\newcommand{\curL}{{\cal L}}
\newcommand{\curO}{{\cal O}}
\newcommand{\vs}{v_{\rm s}}
\newcommand{\phase}{\varphi}
\newcommand{\imag}{{\rm Im}}
\newcommand{\real}{{\rm Re}}
\newcommand{\kb}{k_{\rm B}}
\newcommand{\width}{m}
\newcommand{\curS}{{\cal S}}
\newcommand{\bk}{{\bf k}}
\newcommand{\bq}{{\bf q}}
\newcommand{\bQ}{{\bf Q}}
\newcommand{\bp}{{\bf p}}
\newcommand{\bj}{{\bf j}}
\newcommand{\ba}{{\bf a}}
\newcommand{\bA}{{\bf A}}
\newcommand{\bD}{{\bf D}}
\newcommand{\muh}{\hat{\mu}}
\newcommand{\hh}{\hat{h}}
\newcommand{\Hh}{\hat{H}}
\newcommand{\Gh}{\hat{G}}
\newcommand{\Th}{\hat{T}}
\newcommand{\Gb}{\bar{G}}
\newcommand{\zb}{\bar{z}}
\newcommand{\Deltah}{\hat{\Delta}}
\newcommand{\deltah}{\hat{\delta}}
\newcommand{\bb}{{\bf b}}
\newcommand{\mg}{m}
\newcommand{\sfm}{SF$_{\rm M}$\,}
\newcommand{\boldr}{{\bf r}}
\input{epsf}
\title{Imbalanced Feshbach-resonant Fermi gases}
\author{Leo Radzihovsky}
\address{
Department of Physics, University of Colorado, Boulder,
 Colorado 80309
}
\ead{radzihov@colorado.edu}

\author{Daniel E. Sheehy}
\address{
Department of Physics and Astronomy, Louisiana State
  University, Baton Rouge, Louisiana 70803
}
\ead{sheehy@lsu.edu}

%
\begin{abstract}

  We present an overview of recent developments in 
  species-imbalanced (``polarized'') Feshbach-resonant Fermi gases. We
  summarize the current status of thermodynamics of these systems in
  terms of a phase diagram as a function of the Feshbach resonance
  detuning, polarization and temperature. We review instabilities of
  the s-wave superfluidity across the BEC-BCS crossover to phase
  separation, FFLO states, polarized molecular superfluidity and the
  normal state, driven by the species imbalance.  We discuss different
  models and approximations of this system and compare their
  predictions to current experiments.

\end{abstract}


\maketitle

\section{Introduction}
\label{Introduction}

\subsection{Background and motivation}

Launched by the laboratory achievement of Bose-Einstein condensation
(BEC) in dilute alkali gases~\cite{Anderson95,Davis95}, atomic physics
has undergone nothing short of a revolution.
It has transitioned from a few-body science to the study of many-body
physics featuring thermodynamic phases and phase transitions.
Some of the most exciting experimental developments have included
observations of lattices of quantized vortices in rotating BEC
gases\cite{Matthews99,Madison01,AboShaeer01,Coddington2004}, the Mott
insulator to superfluid transition of bosons in an optical
lattice~\cite{Greiner02,Fisher89,Jaksch98}, the creation of a Fermi
sea of cold fermionic atoms~\cite{DeMarco99}, and the observation of
paired fermionic superfluids that mimic superconductivity in a metal
or superfluidity of
$^4He$~\cite{Regal2004,Zwierlein2004,Kinast2004,Bartenstein2004,Bourdel2004,
  Partridge2005,Chin2004,Zwierlein2005}.

After an early period exploring weakly interacting BEC's and Fermi
gases, recently much attention has focused on strong interactions in
degenerate atomic gases, which can be realized, for example, by
confining the atoms with a periodic optical
potential~\cite{BlochReview}, simulating a solid state crystal.
A complementary way to explore strong
interactions~\cite{GRreview,Giorgini08,KetterleZwierlein} in atomic
gases is through the use of Feshbach resonances
(FR)~\cite{Timmermans99}, abundant in alkali
atoms~\cite{Donley01,Regal03,Bartenstein05}, which allow the
experimental control of the inter-atomic interactions.  This crucial
feature of a FR, its {\it tunability\/}, arises through the Zeeman
shift (detuning, $\delta$) of a diatomic molecular ``closed''-channel
bound state relative to the ``open''-channel atomic continuum. This
allows a degree of control (even in real time) over the strength and
sign of the open-channel atomic
interactions~\cite{agr,Barankov,Altman,Strohmaier}, that is
unprecedented in other (e.g., solid-state) contexts.

Feshbach resonance tunability has led to laboratory realizations and
theoretical proposals for a broad range of quantum many-body
phenomena. The most notable experimental realization is of
fermion-paired $s$-wave resonant superfluidity, that exhibits the
BEC-to-Bardeen-Cooper-Schrieffer (BEC-BCS) crossover between a
coordinate-space-paired condensate of dilute diatomic molecules (for
large negative detuning) to a Fermi-surface momentum-paired BCS regime
of strongly overlapping Cooper pairs (for large positive
detuning)~\cite{Leggett1980,Nozieres,sademelo,Timmermans01,Holland,
  Ohashi,agr,Stajic,Chen2005,Nussinov}.  This remarkable
controllability of the atomic interactions comes along with other
tunable properties, including the background potential (that can be
used to confine an atomic gas to one~\cite{Paredes} or two
dimensions~\cite{Hadzibabic06}, or to impose a periodic optical
potential), and the atom numbers of the trapped atomic species.

The focus of this review is an extremely successful set of experiments
studying species-{\em imbalanced} (``polarized''~\cite{Polnote})
mixtures of two hyperfine states of Feshbach-resonantly interacting
fermionic gases.  The two hyperfine states form a pseudo-spin $1/2$
system, that is thus closely related to electrons in a
metal~\cite{Casalbuoni} or high-density quark matter\cite{Alford}, but
with the advantage of the tunability of the interactions and trapping
potential.

Thus this research effort elucidates the response of interacting
fermions to the frustration introduced by species imbalance, that
despite attractive interactions prevents a complete pairing and drives
phase transitions to a variety of interesting quantum ground states
and thermodynamic phases.
A sufficiently large species imbalance is predicted (and observed) to
destroy superfluid order of resonantly interacting fermions even at
zero temperature. Understanding the resulting strongly interacting
nonsuperfluid ground state is of considerable interest and may shed
light on other strongly correlated states lacking broken symmetry such
as the ``pseudogap'' and marginal-Fermi liquid regimes observed in
high T$_c$ cuprate superconductors~\cite{Timusk,Varma}, heavy-fermion
and Kondo-lattice systems~\cite{Stewart,Coleman}, and the
phenomenology near quantum critical points~\cite{Sachdevbook}.

By identifying the species imbalance, $\Delta N$, with the
magnetization $M$ ($= \Delta N/V$ with $V$ the system volume) and the
two-species chemical potential difference, $\Delta\mu = \mu_\uparrow -
\mu_\downarrow\equiv 2h$ with an effective Zeeman energy $h$, this
research makes a connection to a large body of work on a related
condensed-matter system, namely superconductors under an applied
magnetic field~\cite{Casalbuoni}. Thus, imbalanced Feshbach-resonant
atomic systems allow a natural extension of superconductivity in a
Zeeman field (previously limited to the BCS
regime\cite{Clogston,Sarma}) across the full BEC-BCS
crossover. Furthermore, unlike charged solid-state superconductors,
where Zeeman and orbital fields are necessarily simultaneously induced
by the applied magnetic field, atomic systems are advantageous as they
permit independent control of these two effects, with the former
created by species chemical potential difference, and the latter
induced by the rotation of the atomic cloud~\cite{Ho01,VeilletteRot,Shim}.

Experiments on polarized Fermi gases have been led by the MIT and Rice
groups~\cite{Zwierlein2006,Partridge2006,Shin2006,Partridge2006prl},
that early on observed imbalance-induced phase-separation and explored
theoretical predictions (such as the phase diagram~\cite{SR2006}) for
the BEC-BCS crossover at a finite species
imbalance~\cite{Liu,Bedaque,Mizushima,Carlson,Son,Pao,SR2007,SR2007Comment}.
In such experiments, the system is typically prepared with all the
atoms in one state.  By applying a radio-frequency (RF)
pulse~\cite{Zwierlein2006,Partridge2006}, a controllable number of
atoms can be transferred to the other state, creating an imbalanced
Fermi gas.  The error on the resulting imbalance (measured at the end
of the experiment) is typically a few percent, allowing an accurate
study of the phases of strongly interacting imbalanced fermion gases.

Inspired by the rapid experimental progress, this line of research
exploded with theoretical activity, leading to a wealth of predictions
that continue to be explored in the laboratory. These include an
extension of the $T=0$ microscopic analysis of the detuning-imbalance
phase diagram~\cite{SR2006,SR2007} to finite
temperature~\cite{Parish2007,Chien,Gubbels,Sharma}, proposals for
numerous new phases and transitions, and an analysis of the strongly
interacting normal state appearing at high
imbalance~\cite{Schunck2007,Schirotzek2009,Nascimbene}.  The universal
phenomenology of a
balanced~\cite{Ho2004,Heiselberg2001,Carlson2003,Astrakharchik2004,
  Bulgac2006,Burovski2006} unitary Fermi gas has also been extended to
a finite imbalance~\cite{Chevy,Nikolic2006,Veillette07}. Along with
these theoretical developments, new and powerful experimental probes
(e.g., RF spectroscopy~\cite{Schunck2007}, in-situ density profile
measurements~\cite{Shin2006}, and measurements of collective
modes~\cite{Nascimbene}) have been brought to bear on this rich
system, and in turn have stimulated further theoretical
predictions. An overview of the considerable progress in the current
understanding and of the remaining open questions in these imbalanced
resonant Fermi gas systems is the focus of this review.  Aimed to be
significantly briefer and less comprehensive, it complements
significantly more detailed recent reviews by Gurarie and
Radzihovsky\cite{GRreview}, by Bloch, Dalibard, and
Zwerger~\cite{BlochReview}, by Giorgini, Pitaevskii, and
Stringari~\cite{Giorgini08} and by Ketterle and
Zwierlein~\cite{KetterleZwierlein}. 

The present review provides an overview of the current status of the field
of imbalanced fermion gases, and is considerably broader than our recent 
research article~\cite{SR2007} that only provided the details of our own theoretical
calculations of the $T=0$ mean-field phase diagram. 
In this review, we aim to avoid the details of analyses while discussing the present-day
understanding of  the $T=0$ mean-field theory~\cite{SR2006,SR2007} and the results of finite
temperature mean-field and large-N calculations in 
 Sec.~\ref{SEC:irg}, the nature and stability of strong fluctuations in the FFLO
state in Sec.~\ref{sec:fflo}, the predictions of the local density approximation in 
Sec.~\ref{sec:lda}, and the status of the
most up to date experimental observations on this system in Sec.~\ref{SEC:recentexp}.  
In Sec.~\ref{SEC:concluding}, we conclude with a few remarks about the future directions of 
this exciting subject in the physics of correlated fermions.

%
%

\subsection{Two-body Feshbach-resonant scattering}

Before turning to the analysis of states at finite density, we briefly
review the Feshbach-resonance mediated interaction, that (at finite
density) leads to superfluidity in fermionic atomic gases.  From a
broad perspective, the low-energy atomic scattering and therefore
(through the T-matrix) the effective inter-atomic attraction is
strongly influenced by the presence of a bound (or quasi-bound) state
in the two-atom potential.  More microscopically, Feshbach resonant
scattering in the ``open'' channel takes place when there is an
energetically-nearby ``closed'' channel, that is weakly coupled to it
by the hyperfine interaction.
Crucially, due to a Zeeman splitting between the bound state (closed
channel) and the scattered states (open channel), an external magnetic
field can be used to change the bound-state energy relative to the
continuum of the scattering atoms, and thereby tune the effective
interactions between scattering atoms.

The Feshbach-resonant scattering of the open-channel fermions is
characterized by the s-wave scattering amplitude $f_0(k)$, that by
unitarity and analyticity~\cite{LandauQM} is required to take the form
\be
f_0(k) = \frac{1}{-\as^{-1} + \frac{1}{2} r_0k^2 -ik},
\label{scattamp}
\ee
with $\as$ the scattering length and $r_0$ the effective range
parameter that is negative in the resonant regime~\cite{GRreview}.  As
the closed-channel bound-state energy is tuned through zero by
adjusting an external magnetic field $B$, the scattering length
diverges as
\be
\as =a_{bg}\Big(1-  \frac{B_w}{B-B_0}\Big),
\ee
with $a_{bg}$ the background scattering length, $B_0$ the resonance
position (at which the bound-state energy vanishes) and $B_w$ the
resonance width.

A microscopically faithful way to capture the Feshbach resonance
phenomenology is through the so-called two-channel
model~\cite{Holland,Ohashi,agr}, in which the closed channel molecules
appear explicitly in the Hamiltonian and give the scattering amplitude
in \rfs{scattamp}.
In practice, however, present-day experiments on superfluid Fermi
gases can only access low energies (densities), that can be described
by a simpler, more universal one-channel model to which the
two-channel model reduces in the low-energy, $k |r_0|\ll 1$
limit~\cite{GRreview,Levinsen2005}.  We thus focus on this
experimentally relevant limit, described by the one-channel model
Hamiltonian
\be
H = \sum_{\bk,\sigma}\epsilon_k\ch_{\bk\sigma}^\dagger  \ch_{\bk\sigma}^\phdag 
+ \frac{\fermiint}{\vol}\sum_{\bk\bq\bp} \ch_{\bk\uparrow}^\dagger \ch_{\bp\downarrow}^\dagger 
\ch_{\bk+\bq\downarrow}^\phdag
\ch_{\bp-\bq\uparrow}^\phdag,
\label{eq:singlechannelintropre}
\ee 
for the two species ($\sigma = \uparrow,\downarrow$) of open-channel
fermions created by the anticommuting operator
$\ch_{\bk\sigma}^\dagger$, with the single-particle energy $\epsilon_k
=\hbar^2 k^2/2m$, mass $m$, and volume $\vol$.  The attractive
interactions are parameterized by $\lambda<0$.

Through a standard T-matrix scattering calculation, that gives
$-f_0(k) = \frac{m}{4\pi\hbar^2}T_k$, the pseudo-potential parameter
$\lambda$ can be related to the experimentally determined, magnetic
field dependent~\cite{Regal03,Bartenstein05} scattering length
\be 
\frac{m}{4\pi \as\hbar^2} = \frac{1}{\lambda}+ \frac{1}{\vol}\sum_\bk
 \frac{1}{2\epsilon_k},
\label{eq:scatt}
\ee
where the ultraviolet-divergent second term is regularized by a
microscopic momentum cutoff scale $\Lambda\sim 1/d$ set by the
closed-channel molecular extent $d$. This gives (with $\hbar=1$
throughout)
\begin{eqnarray}
\label{eq:deltalength} 
\as(\lambda)& =& \left({4\pi \over m \lambda} + {2 \Lambda
    \over  \pi} \right)^{-1} \equiv \frac{m}{4\pi}\lambda_R,\\
&=&\frac{m}{4\pi}\frac{\lambda}{1-\lambda/\lambda_c},
\end{eqnarray}
where $\lambda_R$ can be called the renormalized coupling and
$\lambda_c=-{2\pi^2 \over\Lambda m}$ is the critical coupling
$\lambda$ at which the scattering length diverges.  The above relation
allows the definition of the model and therefore a reexpression of
physical observables in terms of the experimentally defined (UV-cutoff
independent) scattering length $\as$.

\subsection{Resonant Fermi gas at finite density: limits of validity}

In contrast to the two-body problem discussed above, a resonant
many-body system at finite density presents a formidable challenge,
that in most cases can only be treated approximately. Although many
uncontrolled (but illuminating) approximations have appeared in the
literature\cite{Timmermans01,Holland,Ohashi,Chen2005} dating back to
the pioneering works studying the BEC-BCS
crossover~\cite{Leggett1980,Nozieres,sademelo}, such one-channel model
studies across the unitary point can only be trusted {\em
  qualitatively}. The reason is that the one-channel model at finite
density $n$ exhibits only a single dimensionless gas parameter
$n\as^3$ (or, equivalently, $\kf \as$, with the Fermi momentum $\kf$),
a measure of the resonant interaction strength, that diverges upon
approach to the Feshbach resonance, where
$|\as|\rightarrow\infty$. Thus, a perturbative expansion in $\kf \as$
is precluded sufficiently close to the resonance and is useful only
far from the resonance, deep in the BCS and BEC limits, where $\kf
|\as|\ll 1$.

Nevertheless, trustworthy treatments of such strongly interacting,
scale-invariant systems are indeed possible, taking a cue from the
studies of critical phenomena~\cite{CriticalPhenomena,Fisher,Wilson},
where analogous challenges have been surmounted almost 40 years
ago. The approach is to ``deform'' the physical model to one that
exhibits a mathematical small parameter, such as the deviation
$\epsilon \equiv d_{uc}-d$ of spatial dimension $d$ from an
upper-critical dimension, $d_{uc}$, above which the behavior is
simple, or a $1/N_f$ expansion about a large number of atom flavors,
$N_f$, with the exactly solvable $N_f\rightarrow\infty$ limit
formalizing and justifying earlier mean-field approximations. There
has been considerable success in applying these field-theoretic
methods to a unitary Fermi
gas\cite{Nikolic2006,Veillette07,Nishida2007,Veillette08}.  These
methods have also been embellished in uncontrolled ``self-consistent''
resummation schemes, that do not always lead to an improved
description~\cite{Haussmann}. From the related renormalization group
point of view, in the solvable vacuum (two particle) limit of the
previous subsection, a Feshbach resonance with $|\as|\rightarrow\infty$
corresponds to a critical point~\cite{Nikolic2006,Veillette07}, that
separates two distinct ``phases'' (vacua at zero density) on the BCS
(positive detuning, $\as < 0$) and BEC (negative detuning, $\as > 0$)
sides of the resonance. At a finite density $n$, this transition
between the corresponding BCS and BEC paired superfluids is converted
to a smoothed crossover, analogous to a paramagnet-ferromagnet
crossover in a magnetic field.

Another useful approach to a controlled theory of the BEC-BCS
crossover is through a narrow Feshbach resonance\cite{agr,GRreview},
described by the two-channel model~\cite{Timmermans99}.  In the
latter, in addition to the gas parameter $\kf \as$, there appears
another dimensionless parameter,
\begin{equation}
  \gamma=\frac{\sqrt{8}}{\pi} \sqrt{\Gamma_0\over \ef}
=\frac{8}{\pi}{1\over \kf |r_0|},
\end{equation}
that is the dimensionless ratio of the width of the resonance,
$\Gamma_0$ (proportional to $B_w$, set by the strength of the Feshbach
resonance coupling, the hybridization amplitude of an atom-pair with a
molecule) to the Fermi energy $\ef$.  This resonance-width parameter
$\gamma$ naturally allows a distinction between  wide ($\gamma\gg 1$)
and narrow ($\gamma\ll 1$) resonances.  Equivalently, these are
contrasted by whether, upon growth near the resonance, the scattering
length $\as(B)$ reaches the effective range $|r_0|$ first (the broad
resonance) or the atom spacing $k_F^{-1}\sim n^{-1/3}$ first (the
narrow resonance).  A quantitatively-accurate description of the full
BEC-BCS crossover, perturbative in $\gamma$, can thus be obtained for
{\it narrow\/} Feshbach resonances, making them attractive from the
theoretical point of view.  The ability to treat narrow resonant
systems perturbatively physically stems from the fact that such an
interaction, although arbitrarily strong at a particular energy, is
confined only to a narrow energy window around the resonance
energy\cite{agr,GRreview}.

In practice most $s$-wave Feshbach resonances studied to date are
broad, $\gamma\gg 1$, with one notable exception discussed in
Ref.~\cite{Strecker}.
However, even for broad resonances (characteristic of a featureless
attractive potential, where due to lack of a large potential barrier
no long-lived resonant state exists at positive energy)
$\gamma$ can be thought of as a mathematical expansion parameter about
the solvable narrow resonance ($\gamma\rightarrow 0$) limit, analogous
to the $1/N_f$ and $\epsilon$ expansions.  Furthermore, because of the
density dependence $\gamma\sim 1/n^{1/3}$, in principle $\gamma$ can
be reduced by working at higher atomic densities.

\subsection{Balanced BEC-BCS superfluid crossover}

Before turning to the main subject of polarized (imbalanced) resonant
Fermi gases, we briefly review the salient points of a {\em balanced}
gas and the associated BEC-BCS crossover.  Since the physical
interaction is everywhere attractive (although the scattering length
changes sign), the ground state of the system is expected to be an
s-wave superfluid for the full range of detuning.  The latter will
transition to a nonsuperfluid thermal state at a sufficiently high
temperature $T_c$ (tunable via the FR), or to a Mott or Bose-glass
insulator when subjected to a sufficiently strong commensurate
periodic potential or quenched disorder\cite{Fisher89}.

The qualitative picture for the superfluid state across the resonance
follows from the aforementioned solvable narrow resonance
limit~\cite{agr,GRreview}, in which the closed channel molecule is
long lived and can therefore be treated as an independent particle
even for positive detuning. For detuning larger than twice the Fermi
energy, closed-channel molecules are too energetically costly and most
of the atoms are in the form of open-channel fermions, forming a
weakly BCS-paired Fermi sea, with an exponentially small molecular
density, induced by the weak Feshbach resonant coupling. The BEC-BCS
crossover initiates as the detuning is lowered below an energy scale
of order of $2\ef$, where a finite density of atoms begins to bind
into Bose-condensed closed-channel molecules, stabilized by the Pauli
principle. The resulting molecular (closed-channel) superfluid
coexists with the strongly-coupled BCS superfluid of (open-channel)
Cooper pairs, that, while symmetry-identical and hybridized with it by
the Feshbach resonant coupling, is physically distinct from it. This
is made particularly vivid in highly anisotropic, one dimensional
traps, where the two distinct (molecular and Cooper-pair) superfluids
can actually decouple due to quantum fluctuations, suppressing the
Feshbach coupling at low energies~\cite{SheehyDecouple}. The crossover
to the BEC superfluid terminates around zero detuning, where the
conversion of open-channel atoms (forming Cooper pairs) into
closed-channel molecules is nearly complete.  In the asymptotic
negative-detuning regime a true bound state appears in the
closed-channel, leading to a positive scattering length and a two-body
repulsion in the open-channel. In between, as the position of the
Feshbach resonance is tuned through zero energy, the system is taken
through (what would at zero density be) a strong unitary scattering
limit, corresponding to a divergent scattering length, that is
nevertheless quantitatively accessible in the narrow resonance limit,
where $\gamma \sim 1/(k_F |r_0|)$ plays the role of a small parameter.

Although the above features of the crossover are no longer expected to
vividly appear in the experimentally relevant broad resonance limit,
it is useful to keep this qualitative picture in mind. The detailed
description of the broad resonance emerges when the Hamiltonian
is treated in the mean-field approximation (although the aforementioned
$\epsilon$- and $1/N_f$- expansions can also be utilized to
systematically correct it). At $T=0$, a standard mean-field analysis
(that is exact in the large $N_f$ limit\cite{Nikolic2006,Veillette07})
of $H_\mu=H-\mu N$ gives the grand-canonical ground state energy
density
\bea
\label{eq:gsesinglechannel}
E_G  &=& -\frac{\Delta^2}{\fermiint} + V^{-1}\sum_k (\xi_k - E_k),\\
&=&-\frac{m}{4\pi \as} \Delta^2 
+ \int \frac{d^3 k}{(2\pi)^3} (\xi_k - E_k + \frac{\Delta^2}{2\epsilon_k}),
\eea
with $\xi_k \equiv \epsilon_k -\mu$ and $E_k=\sqrt{\xi_k^2+\Delta^2}$,
and in the second line $E_G$ was reexpressed in terms of the
scattering length, thereby also leading to a finite pairing integral
(the second term). A minimization of $E_G(\Delta,\mu)$ over the
variational superfluid order parameter $\Delta$ ensures its expression
in terms of the fermion pair
$\Delta=\lambda\langle\ch_\downarrow(\xv)\ch_\uparrow(\xv) \rangle$,
i.e., gives the gap equation, and the chemical potential enforces the
total fermion number equation $n = - \frac{\partial E_G}{\partial \mu}
= V^{-1}\sum_{\bk,\sigma} \langle c_{\bk \sigma}^\dagger c_{\bk
  \sigma}^\phdag \rangle$.

\begin{figure}[tbp]
\epsfxsize=14cm 
\vskip0.25cm \hskip1cm
\centerline{\epsfbox{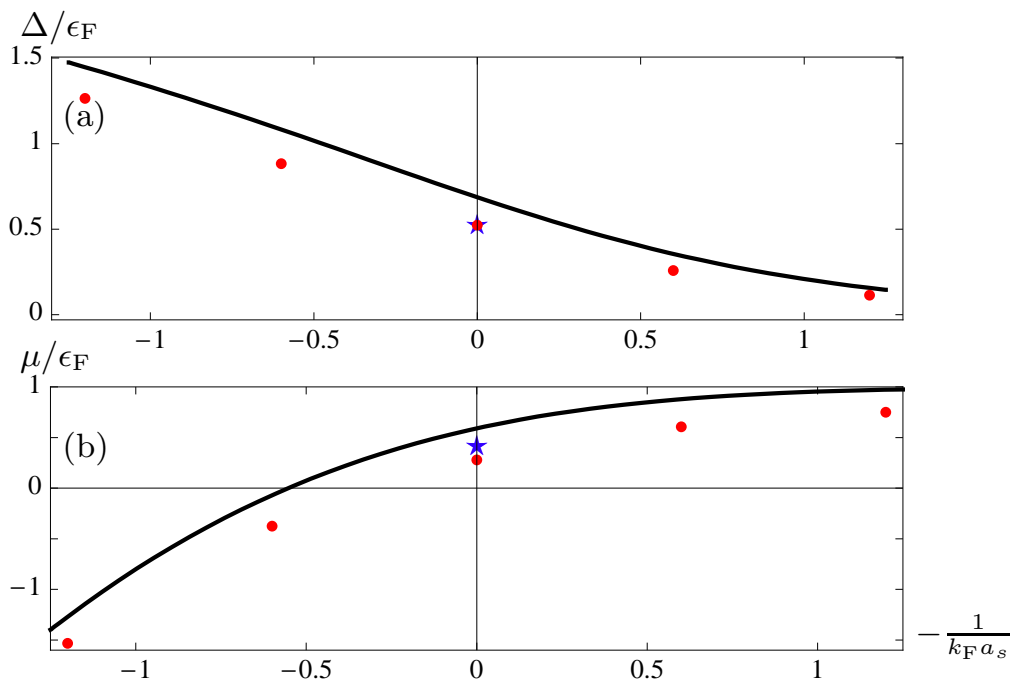}} \vskip.55cm
\caption{Order parameter
  $\Delta$ (panel a) and 
chemical potential $\mu$ (panel b) as a function of $-(\kf\as)^{-1}$, with the BEC
  regime on the left and the BCS regime on the right. The solid black  lines
  are the mean-field results for $\mu$ and $\Delta$, following
  Ref.~\cite{Marini}. The red circles include the ${\cal
    O}(1/N_f)$ corrections~\cite{Veillette07}, evaluated at $N_f=1$.
   The star symbols at
  unitarity are the results of the quantum Monte Carlo calculation of
  Ref.~\cite{Carlson2003}.}
\vskip.55cm
\label{fig:becbcs}
\end{figure}

The integrals in the mean-field ground-state energy
Eq.~(\ref{eq:gsesinglechannel}), and the gap and number equations, can
be expressed in terms of elliptic integrals~\cite{Marini}, and the
simultaneous numerical solution of the gap and number equations yields
the mean-field $\Delta$ and $\mu$ as a function of the detuning
parameter $-\frac{1}{\kf \as}$, plotted as solid lines in
Fig.~\ref{fig:becbcs}.  The red circles include the leading
$1/N_f$ correction, at $N_f \to 1$, following Ref.~\cite{Veillette07},
while the star symbols at the unitary point are quantum Monte Carlo
results~\cite{Carlson2003}.

It is also illuminating to obtain limiting analytical expressions to
the mean-field results. In the BCS regime ($1 < - \frac{1}{\kf \as}$),
$\mu>0$ and $\Delta\ll\mu$, one finds (defining $c =
\frac{m^{3/2}}{\sqrt{2}\pi^2}$, with the three-dimensional density of
states $N(\epsilon) = c\sqrt{\epsilon}$)~\cite{SR2007}:
\be
E_G \approx  -\frac{m}{4\pi \as} \Delta^2 -\frac{8c}{15}\mu^{5/2}
- c\sqrt{\mu} \Delta^2\left(
\frac{1}{2} - \ln \frac{\Delta}{8{\rm e}^{-2} \mu}
\right),
\ee
that is non-analytic in $\Delta$, reflecting the well-known fact that
an arbitrary weak attraction yields a pairing instability.  To leading
order in $\Delta\ll \mu$, this gives the standard BCS expressions for
the gap ($0 = \frac{\partial E_G}{\partial \Delta}$) and number ($n =
- \frac{\partial E_G}{\partial \mu}$) equations
\bea
 \Delta &\approx& \Delta_{\rm BCS} \equiv 
 8 {\rm e}^{-2} \ef \exp\bigg(\frac{\pi}{2\kf \as }\bigg),
\label{eq:deltahsingle}
\\
\mu &\approx&\ef.
\eea
Here, the Fermi energy $\ef$ is defined in terms of the particle
density via $n = \frac{4}{3} c\ef^{3/2}$.  In the asymptotic
negative-detuning BEC regime of $-\frac{1}{\kf \as} < -1$, in which
$\mu<0$ and $|\mu|\gg\Delta$, one instead obtains the analytic Landau
form
\be
E_G \approx  -\frac{m}{4\pi \as} \Delta^2 + \frac{1}{2} c\sqrt{|\mu|} |\Delta|^2
\big( \pi+ \frac{\pi}{32}\frac{\Delta^2}{\mu^2}
\big),
\ee
that leads to the gap and number equations (for $\Delta\ll |\mu|$)
\bea
\mu &\approx&  -\frac{1}{2m\as^2},\\
\Delta &\approx& \ef\sqrt{\frac{16}{3\pi\kf \as}},
\eea
with $\mu$ equal to one-half the two-atom binding energy in the BEC
limit.

As an aside, we note that expressions for the ground state energy, the
chemical potential, and the order parameter (gap) are all well-defined
in the unitary $a_s\rightarrow\infty$ limit, respectively given by
universal forms
\begin{eqnarray}
E_G&=& 0.5906 \frac{3}{5} n\ef,\ \ \ \Delta =0.6864\ef,
\ \ \ \mu =0.5906\ef,
\end{eqnarray}
all expressible in terms of the only energy (or equivalently, density
$n=k_F^3/3\pi^2$) scale $\ef$ of the problem, with the numerical
prefactors approximations (to lowest order in the
$1/N_f$-expansion~\cite{Nikolic2006,Veillette07}) for what are
expected to be universal numbers characteristic of a unitary Fermi
gas.

The above analysis can be generalized to finite temperature, giving
the transition temperature $T_c$ to the normal state, at unitarity
also scaling with $\ef$ (but which becomes inaccurate in the deep-BEC
limit~\cite{Nozieres}). A finite temperature also introduces an
additional energy scale, at unitarity leading to the generalization of
the above numerical prefactors to dimensionless universal scaling
functions of $\ef/T$.

\section{Imbalanced resonant Fermi gas}
\label{SEC:irg}
\subsection{Model}
With this background in mind the {\em imbalance} in the number of the
two species is incorporated by introducing two distinct chemical
potentials, $\mu_\sigma=(\mu_\uparrow,\mu_\downarrow)$ to impose
separately conserved fermion numbers $N_\sigma =
(N_\uparrow,N_\downarrow)$, or equivalently total fermion number
$N=N_\uparrow + N_\downarrow$ and species imbalance $\Delta N =
N_\uparrow - N_\downarrow$.  The corresponding grand-canonical
Hamiltonian $H_{\mu_\sigma}=H-\mu N - h\Delta N$ is
\be
\curH = \sum_{\bk,\sigma}(\epsilon_k-\mu_\sigma)\ch_{\bk\sigma}^\dagger  \ch_{\bk\sigma}^\phdag 
+ \frac{\fermiint}{\vol}\sum_{\bk\bq\bp} \ch_{\bk\uparrow}^\dagger \ch_{\bp\downarrow}^\dagger 
\ch_{\bk+\bq\downarrow}^\phdag
\ch_{\bp-\bq\uparrow}^\phdag,
\label{eq:singlechannelintro}
\ee
with $\mu_\uparrow = \mu + h$ and $\mu_\downarrow = \mu - h$ related
to the total-number chemical potential $\mu$ and species-imbalance
chemical potential $h$ (an effective Zeeman energy).  The imbalanced
resonant Fermi gas thermodynamics as a function of $N,\Delta N, T,
\as$, i.e., the extension of the BEC-BCS crossover to a finite
imbalance can then be computed by a variety of theoretical techniques,
including quantum Monte Carlo~\cite{Carlson,Pilati}, mean-field
theory~\cite{SR2007,SR2007Comment,Pao,SR2007,Parish2007}, the
large-$N_f$ (fermion
flavor)~\cite{Nikolic2006,Veillette07,Veillette08} and $\epsilon$
expansions~\cite{Nishida2007}.  By studying
Eq.~(\ref{eq:singlechannelintro}) we have already specialized to the
wide Feshbach resonance limit. However, this problem can also be
studied in a controlled narrow resonance regime by working with a
two-channel model as done in Refs.~\cite{SR2006,Parish2007}.

\subsection{Analysis at zero temperature}
We now outline the resulting picture (that is in qualitative agreement
with experiments~\cite{Zwierlein2006,Partridge2006}) using standard
mean-field theory~\cite{SR2007} in the broad-resonance limit.  At zero
temperature this amounts to a minimization of the mean-field
ground-state energy $E_G$ with respect to the variational parameters
(defined below) at fixed $\mu$ and $h$.  The derivatives with respect
to $\mu$ and $h$
\bea
\label{eq:num}
\frac{N}{V} &=&-\frac{\partial E_G}{\partial\mu},\ \ \ 
\label{eq:pol}
\frac{\Delta N}{V} =-\frac{\partial E_G}{\partial h}, 
\eea
then respectively give the total atom number,
$N=N_\uparrow+N_\downarrow$ and species imbalance, $\Delta
N=N_\uparrow-N_\downarrow$, imposed in cold-atom experiments. Our
variational ansatz assumes pairing of the form
\be
\label{eq:deltavx}
\Delta(\bx) = \Delta_\bQ {\rm e}^{i\bQ\cdot \bx},
\ee
with $\Delta_\bQ$ and $\bQ$ variational parameters (with $\bQ\neq 0$ to
allow for the possibility of the simplest
Fulde-Ferrell-Larkin-Ovchinnikov (FFLO)~\cite{FF,LO} type state).
With this, one can derive the ground-state energy density
$E_G(\Delta_\bQ,\bQ,\mu,h)$ following Ref.~\cite{SR2007}
\bea
&&E_G = -\frac{m}{4\pi \as}|\Delta_\bQ|^2 + 
 \int \frac{d^3 k}{(2\pi)^3} ( \xi_k + \frac{Q^2}{8m} - E_k  + \frac{\Delta_\bQ^2}{2\epsilon_k}) \label{eq:havgsingle}
 \\ 
&& 
 \qquad +  \int \frac{d^3 k}{(2\pi)^3} \big[
E_{\bk +} \Theta(-E_{\bk +})+ E_{\bk -} \Theta(-E_{\bk -})
\big],
\eea
with $\Theta(x)$ the Heaviside step function and where we defined
\bea
E_k &\equiv& \Big[\big(\xi_k + \frac{Q^2}{8m}\big)^2 +\Delta_\bQ^2\Big]^{1/2},
\label{eq:excitationenergy}
\\
E_{\bk\pm} &\equiv& E_k \mp\big(h - \frac{\bk \cdot \bQ}{2m}\big),
\label{ekuparrow}
\label{ekdownarrow}
\eea
which can be solved for the stationary $\Delta_\bQ$ and $\bQ$ by
supplementing the gap equation with the condition
\be
\frac{\partial E_G}{\partial Q} = 0. 
\ee
Although the mean-field approximation is expected to be quantitatively
inaccurate in the strongly-interacting unitary regime, it does provide
a consistent qualitative picture and a natural starting point for more
accurate theoretical methods.

An important point to keep in mind, for a proper analysis of
$E_G(\Delta_\bQ,\bQ,\mu,h)$, is that not every simultaneous solution of
the gap and number equations corresponds to a physical ground state of
the system; the key additional criterion is that the solution $\Delta_\bQ$
must also be a {\em minimum} of $E_G$ at fixed $\mu,h$. The
verification that an extremum solution is indeed a minimum is
particularly essential when (as is the case for a polarized Fermi gas)
there is a possibility of a first-order transition, where $E_G$
exhibits local maxima and saddle points that separate its local
minima. By working instead directly with the gap equation and imposing
fixed $N$ and $\Delta N$ from the start, several authors have been
lead to qualitatively and/or quantitatively incorrect phase
diagrams~\cite{Pao,Iskin,Chien}, mistakenly associating local minima,
saddle-points and/or even maxima with a ground state, and
misidentifying first-order phase boundaries with the spinodals, as
discussed in Refs.~\cite{SR2007Comment,LamacraftMarchetti}.

Therefore, in our experience the safest path to the correct phase
diagram is to work in the grand-canonical ensemble, use the global
minimum of the free energy to map out the phase diagram at fixed
$\mu,h$, and only then constrain it to the imposed total number and
imbalance densities; the in-satisfiability of this final constraint is
a signature of phase separation, resolved by an inhomogeneous
coexistence of two phases in proportions imposed by the two density
constraints.
\begin{figure}[tbp]
\epsfxsize=9cm 
\centerline{\epsfbox{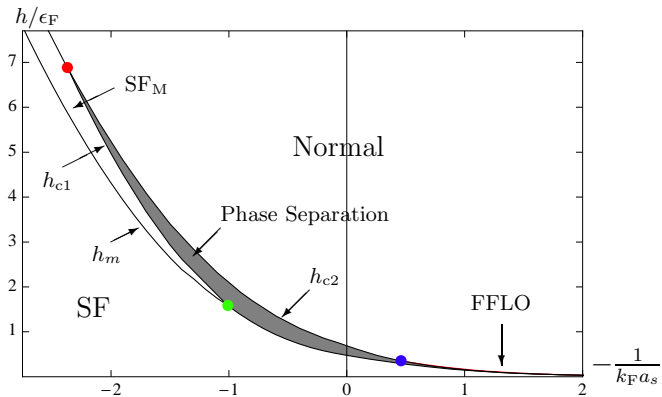}} \vskip-.25cm
\caption{Mean-field phase diagram of imbalanced Fermi gases, as a
  function of inverse scattering length and chemical potential
  difference (Zeeman energy) $h$, showing regimes of an imbalanced
  magnetized superfluid (SF$_{\rm M}$) bounded by $h_m$ and $h_{{\rm
      c}1}$, and a strongly-interacting normal Fermi liquid. 
The region denoted $SF$ is the balanced BEC-BCS crossover superfluid
phase.  
The central shaded region is a regime of phase separation, where the gas
  inhomogeneously coexists between two phases.  The FFLO region is
  indicated by an arrow and bounded, to the right of the blue point by
  $h_{\rm FFLO}$ and $h_{{\rm c}2}$; however it is too small to be
  seen on this scale. A zoom-in of this region is shown in
  Fig.~\ref{fig:phasediagram_fflo}.  The red point at $(-2.37,6.89)$
  is a quantum tricritical point (QTP) separating the first- and
  second-order transitions from the SF$_{\rm M}$ to the polarized
  Normal phase. The green point at $1/(k_F a)\approx 1.01, h\approx
  1.5$ is a critical end-point.   }
\label{fig:phasediagram_h}\vspace{-.1cm}
\end{figure}

\subsection{Phase diagram}

\subsubsection{Zero temperature}

\begin{figure}[tbp]
\epsfxsize=9cm 
\centerline{\epsfbox{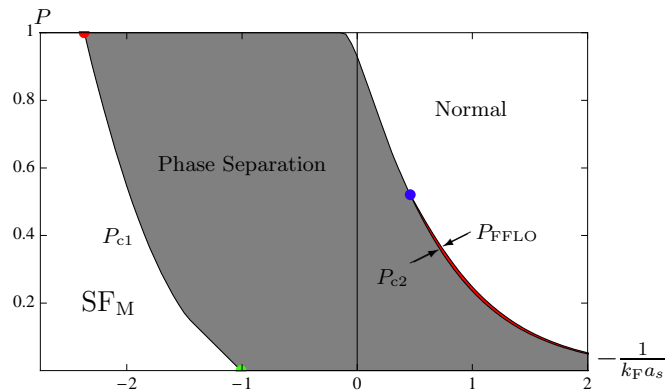}} \vskip-.25cm
\caption{Mean-field phase diagram of imbalanced Fermi gases, as a
  function of inverse scattering length and imbalance $P$, showing
  regimes of magnetized (imbalanced) superfluid (SF$_{\rm M}$), FFLO
  (shaded red, bounded by $P_{\rm FFLO}$ and $P_{{\rm c}2}$) and
  normal Fermi liquid.  The central shaded region is a regime of phase
  separation, and the red point at $(-2.37,1)$ is a quantum
  tricritical point (QTP). The FFLO phase is only stable to the right
  of the blue point.  The green point at $1/(k_F a)\approx 1.01, P=0$
  is a critical end-point.}
\label{fig:phasediagram}
\end{figure}

\begin{figure}[tbp]
\epsfxsize=9cm 
\centerline{\epsfbox{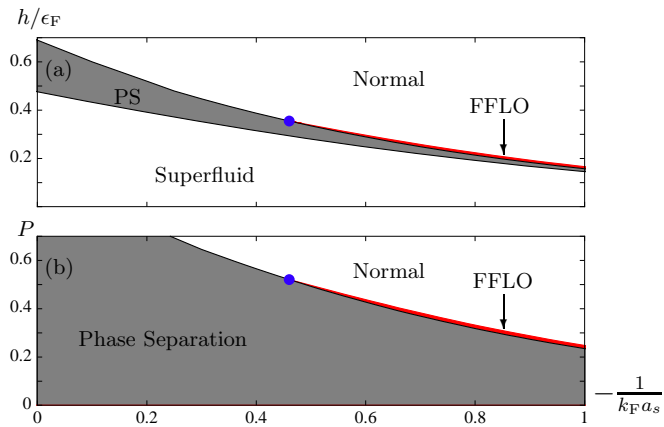}} \vskip-.05cm
\caption{Panel a is a zoom-in of the phase diagram of
  Fig.~\ref{fig:phasediagram_h} at fixed $h$ and inverse scattering
  length.  Panel b is a zoom-in of the phase diagram of
  Fig.~\ref{fig:phasediagram} at fixed $P$ and inverse scattering
  length.  The horizontal axes are the same in the two panels.  In
  each, the FFLO phase is only stable to the right of the blue point
  (as in Figs.~\ref{fig:phasediagram_h} and \ref{fig:phasediagram}).
}
\label{fig:phasediagram_fflo}
\end{figure}

With these cautionary remarks in mind, as detailed in
Ref.\cite{SR2007} $E_G(\Delta_\bQ,\bQ,\mu,h)$ can be analyzed analytically
in a number of asymptotic regimes and numerically throughout. The
result of its minimization directly leads to the zero-temperature
phase diagram~\cite{SR2007} at fixed density and chemical potential
difference in Fig.~\ref{fig:phasediagram_h} and at fixed density and
imbalance in Fig.~\ref{fig:phasediagram}.  The BCS regime of both
figures is shown in Fig.~\ref{fig:phasediagram_fflo}.

The qualitative features of the phase diagram can be readily
understood. At sufficiently small Zeeman energy $h$ the fully-paired
$P=0$ superfluid state (that has $\bQ= 0$, with $\Delta_{\bQ =0}=\Delta=\Delta_{BEC-BCS}$) is stable
across the full BEC-BCS crossover. As first shown by
Sarma~\cite{Sarma}, in the BCS regime the paired state cannot
accommodate any polarization (except for the $Q\neq 0$ FFLO state,
which we shall neglect for the moment).  This follows because the BCS
s-wave paired state involves pairing at the Fermi surface and is fully
gapped. However, in the presence of a population imbalance, the
non-interacting Fermi surfaces have different volumes, determined by
the Fermi momenta $\kfup$ and $\kfdown$, disrupting pairing.  With
increasing $h$ and at fixed $\mu$, then, the BCS state undergoes a
first-order transition to an unpaired magnetized normal state.  At
imposed average density (atom number $N$) and fixed $h$, this leads to
phase separation for $h>h_{c1}$, as shown in
Fig.~\ref{fig:phasediagram_h}.  Since the polarization only becomes
nonzero for $h>h_{c1}$ in the BCS regime, it is clear that, at fixed
polarization $P$, one finds phase separation at arbitrarily small $P$,
as shown in Fig.~\ref{fig:phasediagram}.  Thus, an addition of excess
spin-$\uparrow$ atoms to the paired BCS state leads to their spatial
phase separation from the paired gas.
In contrast, as illustrated in Fig.~\ref{fig:phasediagram}, in the
deep BEC regime the imposed polarization is accommodated by a
transition of the $P=0$ molecular superfluid to a polarized
superfluid, SF$_{\rm M}$, that is a {\it homogeneous\/} ground state
of condensed molecules and a single-species (fully-polarized) Fermi
gas.

The low-$h$ stability of the BEC-BCS superfluid is limited by the
lower-critical Zeeman field, that in the weakly-paired BCS regime ($-1/\kf\as
\gg 1$) is of the order $h_{c1}\sim \Delta_{\rm BCS}/\sqrt{2}$,
exponentially small in $1/\kf \as$ (this is the Clogston
limit~\cite{Clogston}).
We define $h_{c1}$ to be the Zeeman field above which the system phase
separates.
With decreasing detuning, it grows to $h_{c1}\sim \ef$ around the
unitary crossover regime ($-1<-1/\kf\as < 1$).
Deep in the BEC  regime ($1/\kf\as \ll -1$), the stability of the
BEC-BCS superfluid is instead controlled by the molecular binding
energy $h_m\sim E_b\sim 1/m\as^2$.

In the opposite limit of large Zeeman field, $h > h_{c2}$ (at fixed
$N$), a normal (unpaired) state is clearly stable, with $h_{c2}$
exponentially small (but larger than $h_{c1}$) in the BCS limit.  With
decreasing detuning, $h_{c2}$ grows through $\sim\ef$ in the crossover
regime and increases further with the binding energy (while satisfying
$h_m<h_{c1}<h_{c2}$) deeper in the BEC regime.  The high field-induced
depaired state is a partially (two Fermi surfaces) or fully (single
Fermi surface) polarized normal Fermi liquid, $\Delta=0, P\neq 0$,
respectively depending on the Zeeman field strength $h$ and the
interaction strength $1/k_F a$.  Deep in the BEC regime this normal
state is fully polarized, characterized by a single-species Fermi
surface.

In the deep BCS regime, the interactions are weak and the large $h$
depaired state is clearly a weakly renormalized polarized Fermi
liquid, while in the deep BEC regime the only depaired state is fully
polarized ($P=1$), that is, therefore, noninteracting.  In contrast,
in the crossover regime ($-1<-1/\kf\as < 1$) the interactions are
resonant and strong, and therefore cannot be treated
perturbatively. Nevertheless one can explicitly demonstrate using
many-body $1/N_f$-expansion techniques, that (earlier suggestions
based on RF spectroscopy notwithstanding~\cite{Schunck2007}) this
depaired state remains a normal Fermi liquid, albeit a strongly
renormalized one~\cite{Veillette08}. For example, at unitarity and
$P\rightarrow 1^-$ it is characterized by a suppressed quasi-particle
residue $Z = 0.47$, an enhanced effective mass $m_* \approx 1.8 m$,
and a downward chemical potential shift $\delta\mu_\downarrow
=-1.46\ef$.

In the presence of a population imbalance, another possible ground state is the FFLO
phase, in which atoms pair at a nonzero center of mass momentum
$Q\simeq \kfup - \kfdown$ to accommodate the spin-$\uparrow$ and
spin-$\downarrow$ Fermi-surface mismatch.  Within mean-field theory,
and assuming the simplest FFLO-type state, Eq.~(\ref{eq:deltavx}), in
the BCS regime the above-mentioned first-order transition with
increasing $h$ is into the FFLO phase.  With further increase in $h$
the FFLO phase undergoes a continuous transition into the normal phase
yielding the phase diagram at fixed $h$ and fixed
density shown in Fig.~\ref{fig:phasediagram_fflo}a, and the phase
diagram at fixed $P$ and fixed density shown in
Fig.~\ref{fig:phasediagram_fflo}b.  We observe that the FFLO phase is
favored only for a thin sliver of $h$ or $P$, and for sufficiently
{\it weak\/} interactions $\kf |\as| < 2$ in the BCS regime.  In
Sec.\ref{sec:fflo}, we discuss FFLO-type states  in more detail and
the strong effect of fluctuations on them.

Turning to the BEC regime, as illustrated in
Fig.~\ref{fig:phasediagram_h}, at large negative detuning ($-1/\kf\as
< -1$) and at intermediate Zeeman fields, $h_{m}<h<h_{c1}$, an imposed
species imbalance can be accommodated by a homogeneous polarized
(imbalanced) superfluid state, SF$_{\rm M}$ (that we originally dubbed
``magnetized superfluid''\cite{SR2006})\cite{commentSarma}.  This is
not surprising as in this regime of strong attractive interactions
pairing takes place in real space and a nonzero imbalance corresponds
to the addition of a low density of excess fermions to a fluid of
tightly-bound molecular bosons. As discussed in
Ref.~\cite{bulgac2007}, this is analogous to the well-studied problem
of $^3$He-$^4$He mixtures (and, generally, boson-fermion
mixtures~\cite{Viverit,Taylor07}), with fermionic $^3$He atoms
corresponding to the excess spin-$\uparrow$ fermions and known to
exhibit a homogeneous phase (miscible in $^4$He) below a concentration
of about $7\%$ (at $T = 0$)~\cite{Graf67,Schermer}.

In the simplest approximation, the SF$_{\rm M}$ phase is characterized by a ground
state that is the product of a fully polarized majority Fermi sea and a
diatomic molecular condensate.
This phase is therefore quite novel, as it is a
hybrid state displaying strongly-paired superfluidity and gapless
fermionic excitations around a single polarized Fermi surface, and is
thereby expected to display features (e.g., linear low-temperature
heat capacity, superflow, etc.)  characteristic to both.  The SF$_{\rm
  M}$ is separated from the balanced $P=0$ superfluid by a
continuous transition at $h=h_m$, the latter set by the binding
energy.

However, as illustrated in Figs.~\ref{fig:phasediagram_h} and
\ref{fig:phasediagram}, over the large shaded portion of the phase
diagram, at intermediate detuning and Zeeman field (and polarization),
it is not possible to accommodate a species number imbalance with a
homogeneous, single component global minimum of $E_G$.  This happens
whenever two competing ground-state minima become degenerate, and are
characterized by two distinct densities and polarizations for the same
critical values of chemical potentials.  Thus, as is standard for
first-order transitions, it is not possible for the system to achieve
a homogeneous state with intermediate values of the imposed
densities~\cite{SR2006,SR2007}. For the corresponding range of
parameters in the shaded region the system thus phase
separates~\cite{Bedaque} into two coexisting degenerate ground states
with different densities in chemical equilibrium, such that the total
imposed number and polarization constraints are satisfied.  The
resulting phase-separated state can be explicitly accounted for by
generalizing the ground-state ansatz to include the possibility of
such an inhomogeneous mixture~\cite{SR2007}.

One feature of the above phase diagram that has attracted recent
attention is the quantum tricritical~\cite{Griffiths70} point
(QTP)~\cite{Parish2007,SR2007}, appearing at $-1/(\kf \as)\simeq
-2.37$ in the mean-field approximation and denoted as a red point
Fig.~\ref{fig:phasediagram}.  It separates a first-order (to the
right) and second-order (to the left) $P$-increasing transition out of
the polarized superfluid, SF$_{\rm M}$.
Upon decreasing $P$ from unity, the QTP separates two possible fates
of an added spin-down fermion immersed in the polarized Fermi sea: To
the left of the QTP, the magnetized superfluid phase appears, as added
spin-down atoms form tightly-bound pairs with spin-up atoms, miscible
with the remaining spin-up Fermi sea.  In contrast, to the right of
the QTP, the added spin-$\downarrow$ atoms will form molecular pairs,
that are immiscible with the remaining majority Fermi sea (leading to
phase separation to the right of the QTP).  The third possibility for
the added spin-$\downarrow$ atoms is to form a spin-$\downarrow$ Fermi
sea. This is realized to the right in the phase diagram, beyond
$-\frac{1}{\kf \as} \simeq -0.1$, as seen in
Fig.~\ref{fig:phasediagram}, where the phase separation region meets
the normal Fermi liquid region at $P\to 1$.

A similarly interesting region of the phase diagram is the point
(marked green in Figs.~\ref{fig:phasediagram_h} and
\ref{fig:phasediagram}) at which the SF$_{\rm M}$ state emerges (a
critical end point, at which a line of second order transitions
terminates at a first-order transition).  This point is located on the
$P=0$ axis, at $-1/(\kf \as)\simeq -1.01$ in the mean-field
approximation and separates first- and second-order transitions, with
increasing $P$, out of the fully paired, balanced BEC-BCS superfluid.

The above outlined mean-field phase diagram is in qualitative
agreement with recent QMC calculations~\cite{Pilati}, with the
quantitative discrepancies primarily traceable to the neglect, within
mean field theory, of normal state interactions~\cite{Recati2008}.
The latter lower the normal state free-energy and corresponding $P_{c2}$
below their mean-field values.  Indeed, recent
experiments~\cite{ShinNature,ShinPRL2008} have observed the $P_{c2}$
phase boundary (but not the FFLO phase) in qualitative agreement with
Fig.~\ref{fig:phasediagram}, finding, at unitarity, $P_{c2} \simeq 0.3-0.36$ (Fig.2 in
Ref.~\cite{ShinPRL2008}) significantly below its mean-field prediction
of $P^{mf}_{{\rm c}2} \simeq 0.93$, but quite close to
the result of the large-$N$ expansion~\cite{Veillette07} of $P^{large
  N}_{{\rm c}2} \simeq 0.302$ and the QMC result~\cite{Pilati} of
$P^{QMC}_{{\rm c}2} \simeq 0.4$.

However, experiments have not yet detected the $P_{{\rm c}1}$ boundary
separating the SF$_{\rm M}$ and phase separation regimes, although the
finite-temperature SF$_{\rm M}$ phase has been observed; mean-field
theory and QMC predict similar values for $P_{{\rm c}1}$.
To compare, we note that the tricritical point at $P=1$ occurs at
$(\kf \as)^{-1} = 2.37$ in MFT and $(\kf \as)^{-1} = 2.14$ within
QMC~\cite{Pilati}.  Additionally, the onset of the SF$_{\rm M}$ phase
occurs at $(\kf \as)^{-1} =1.01$ in MFT and at $(\kf \as)^{-1} =0.53$
within QMC~\cite{Pilati}.

One regime in which the problem simplifies (while still exhibiting
strong-correlation effects) is the extreme polarized limit
$P\rightarrow 1^-$ of a small density of spin-$\downarrow$ atoms in a
Fermi sea of spin-$\uparrow$ atoms, with the limiting case of one
``impurity'' spin-$\downarrow$ atom immersed in the majority
(spin-$\uparrow$) Fermi sea.  This problem has been studied using 
variational wavefunction~\cite{Chevy2,combescot2007} and Monte
Carlo~\cite{lobo2006} techniques, finding that one spin-$\downarrow$
Fermi atom interacts strongly with the spin-$\uparrow$ Fermi sea
(forming a ``polaron''), as reflected in values of the quasiparticle
residue $Z$, effective mass $m_*$ and chemical potential
$\delta\mu_\downarrow$ that deviate from that of free fermions.  These
polarons have been subsequently observed in experiments studying the
$P\to 1^-$ limit, obtaining (at unitarity) $Z \approx
0.39$~\cite{Schirotzek2009}, $m_* \approx 1.17m$~\cite{Nascimbene},
and $\delta\mu_\downarrow\approx-0.64\ef$~\cite{Schirotzek2009},
values that are in good agreement with QMC and variational
theories. Upon increasing the strength of the attractive interactions
into the BEC regime, around $1/k_F a\approx 1.27$ a phase transition
is predicted~\cite{Punk2009} from a polaron-like state of the minority
atom dressed by the majority Fermi sea (characterized by a nonzero
fermionic pole with a finite residue $Z<1$) to a bosonic molecule
dressed by the (majority) polarized Fermi sea and characterized by a
vanishing residue of the fermionic pole ($Z\to 0$). The properties of
the latter state have been recently analyzed by a molecular
variational wavefunction~\cite{Punk2009,Mora}, thereby constructing a
phase diagram analogous to Fig.~\ref{fig:phasediagram_h}.

In addition to the basic (and well accepted) qualitative features of
the imbalanced resonant paired superfluid, summarized by
Figs.~\ref{fig:phasediagram_h},\ref{fig:phasediagram}, and
\ref{fig:phasediagram_fflo}, a number of additional phases have been
suggested that are not captured by the simplest mean-field ansatz.
These include breached pair phases~\cite{Liu}, deformed Fermi surface
(nematic) phases~\cite{Oganesyan,Sedrakian}, possible further
instabilities of the fully-polarized Fermi sea of the $SF_{\rm M}$
phase~\cite{bulgac2007}, and a number of more exotic
fluctuation-driven FFLO-type phases that we discuss in
Sec.~\ref{sec:fflo}.

\subsubsection{Non-zero temperature}
\label{finiteT}

\begin{figure}[tbp]
\epsfxsize=12cm 
\centerline{\epsfbox{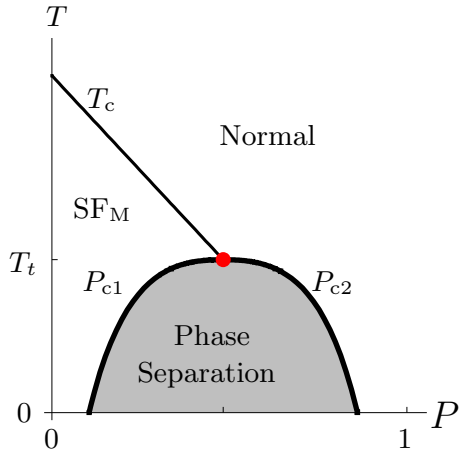}} \vskip-.25cm
\caption{A schematic phase diagram for fixed detuning, illustrating an
  extension of the ground-state phase diagram,
  Fig.~\ref{fig:phasediagram}, to finite $T$, supported by theoretical
  calculations in the narrow-resonance regime~\cite{Parish2007}.  The
  red dot is a tricritical point.  For $-\frac{1}{\kf\as}>
  -2.37$, to the right of the QTP, the phase boundaries $P_{{\rm c}1}$
  and $P_{{\rm c}2}$ extend to finite $T$, merging at a tricritical
  point.  In the regime where the FFLO can be stable, one expects a
  sliver of FFLO phase along the $P_{{\rm c}2}$ boundary. The
  schematic phase diagram depicts a putative regime of detuning, where
  at $T\to 0$, $P_{{\rm c}1}>0$ and $P_{{\rm c}2}<1$.  Such a regime
  does not appear within the mean-field approximation, but is found in
  QMC calculations~\cite{Pilati}. }
\label{fig:phasediagramT}
\end{figure}

Soon after the zero-temperature phase diagram in
Fig.~\ref{fig:phasediagram} was established~\cite{SR2006,SR2007}, the
analysis was generalized to a finite
temperature\cite{Parish2007,Chien,Gubbels,Sharma}, as illustrated in
Fig.\ref{fig:phasediagramT}. As usual, the finite-temperature
thermodynamics and the corresponding phase diagram emerges from the
minimization of the free energy $F$, that for the imbalanced resonant
Fermi gas (because of a number of possible extrema) must be done with
care similar to that of the zero-temperature case.  While the $T=0$
phase diagram is largely established (despite an ongoing search for
the FFLO phase, remaining questions about its extent to smaller
imbalance, and the need for more detailed studies of the $SF_M$), the
status of the finite-$T$ phase diagram in the strongly-interacting
regime is more uncertain. The basic reasons for this can be traced to
the original analysis by Nozieres and Schmitt-Rink~\cite{Nozieres} on
the finite temperature balanced paired Fermi gas and its BEC-BCS
crossover.  While the simple BEC-BCS variational mean-field
wavefunction gives a reasonable approximation for $T=0$ properties
across the phase diagram, for the gas at finite temperature one must
include the effect of thermally excited molecular pairs 
and fermionic single-particle excitations
to obtain a reasonable $\tc$. Because these excitations are expected
to be strongly interacting around the unitary regime (FR point) a
quantitatively trustworthy description remains a challenging open
problem for both balanced and imbalanced gases.

However, quantitative predictions for phase boundaries
notwithstanding, it is likely that the form of the finite-T phase
diagram, predicted by Ref.~\cite{Parish2007} via a controlled narrow
resonance two-channel model
analysis~\cite{agr,GRreview,SR2006,SR2007} is qualitatively correct.  
The finite-$T$ phase diagram, illustrated Fig.~\ref{fig:phasediagramT} is reminiscent of
the phase diagram of $^3$He-$^4$He mixtures.

A number of new ingredients emerge at finite temperature. Because (in
contrast to the fully-gapped BEC-BCS paired ground state) ``spin''-ful
quasi-particles can carry a finite imbalance, at finite temperature a
chemical potential imbalance $h$ always induces a nonzero species
polarization. Thus at finite $T$ the BEC-BCS singlet superfluid state
will no longer be distinguished from the SF$_{\rm M}$ by a vanishing
polarization. Therefore at nonzero temperature $h_m$ is a crossover,
that at low temperature is expected to remain relatively sharp.  On
the BCS side of the resonance at low $T$ the first-order transition
out of the paired BCS superfluid is expected to survive, though likely
to reduce in strength.

Another finite temperature effect is the migration of the
zero-temperature tricritical point of Fig.~\ref{fig:phasediagram} into
the $P<1$ regime~\cite{Parish2007,Gubbels,SheehyTri}, as illustrated
in Fig.~\ref{fig:phasediagramT}.  Finally, as we will discuss below, a
strikingly more significant qualitative effect of thermal fluctuations
is predicted to arise in the putative unidirectional FFLO
states\cite{LeoAshvin}. The enhanced role of fluctuations in such
periodically paired superfluids is associated with underlying
rotational and translational invariance that is spontaneously
broken. As we will discuss in more detail in the next section,
fluctuations lead to universal power-law density correlations in the
LO state, and a melting of the translational order (that can liberate
fractional charge dislocations and vortices) and induce quantum
nematic phases.

\section{Fulde-Ferrell-Larkin-Ovchinnikov superfluid}
\label{sec:fflo}
One of the more interesting alternatives to the resolution of the
species-imbalance imposed frustration is the
Fulde-Ferrell-Larkin-Ovchinnikov (FFLO)\cite{FF,LO} superfluid. In
this state the Fermi surface mismatch is accommodated by finite
momentum pairing, that in its most generic form is a paired
supersolid, that is somewhat of a holy grail in condensed matter
physics. First proposed more than 45 years ago, this enigmatic paired
state has been explored extensively in the context of solid state
superconductors and high-energy physics
systems~\cite{Casalbuoni,Alford}.  However, with the exception of
recent promising experiments on CeCoIn$_5$\cite{Radovan03,Bianchi03},
the FFLO state has so far eluded definitive observations.

\subsection{Theoretical description}

The existence of the FFLO state is most easily detected,
theoretically, through a transition from the polarized normal state
upon reduction of the chemical potential difference, $h$ (or
equivalently, by reduction of the species imbalance). Assuming (and
verifying a posteriori) a continuous (or {\em weakly} first-order)
N-FFLO transition, a perturbative treatment via a Landau expansion in
the pairing order parameter $\Delta_\Qv$, defined by
$\Delta(\xv)=\sum_{\Qv_n}\Delta_{\Qv_n} e^{i\Qv_n\cdot\xv}$, is
justified. Standard analysis gives Landau free energy density
\begin{eqnarray}
  \cH_{GL}&=&\sum_\qv\Delta^*_\qv\eps_Q\Delta_\qv 
+\sum_{\qv_1,\qv_2,\qv_3}
v_{\qv_1,\qv_2,\qv_3,\qv_4}\Delta^*_{\qv_1}\Delta_{\qv_2}
\Delta^*_{\qv_3}\Delta_{\qv_4}+\ldots,
\label{cHgl_form}
\end{eqnarray}
where translational invariance constrains $\eps_Q$ to be a function of
the magnitude of $\qv$ only, $\qv_4=\qv_1-\qv_2+\qv_3$, $\Delta_\qv$ is
a Fourier transform of $\Delta(\xv)$, and
\begin{eqnarray}
\eps_Q\approx \frac{3n}{4\ef}\left[-1
+\frac{Q^2}{2\gamma\mu} 
+ \frac{1}{2}\ln\frac{v_F^2Q^2-4h^2}{\Delta_{\rm BCS}^2}
+ \frac{h}{v_FQ}\ln\frac{v_F Q+2h}{v_F Q-2h}\right],
\label{epsSR}
\end{eqnarray}
obtained either by a direct Landau expansion\cite{LO} or via expansion
of the Bogoliubov-de Gennes (BdG) equation\cite{SR2007} to quadratic
order in $\Delta_\qv$.  Simple analysis shows that $\eps_Q$ has a
minimum at $|\Qv|\approx 1.81\Delta_{\rm BCS}/\hbar v_F \approx
0.58/\xi_0$, infinitely degenerate with respect to the detailed nature
of the FFLO state, thereby leaving the quartic interactions to lift
this degeneracy, i.e., to select the set of $\Qv$'s defining a
specific type of the FFLO state.  As first demonstrated by LO, just
below $h_{c2}$ it is the $\pm Q$ periodic (cosine) state that is
selected.  

However, the above treatment cannot be justifiably extended far below
$h_{c2}$, down to the transition to the uniform fully-paired
superfluid state near $h_{c1}$. There, the order parameter is no
longer small and therefore must be treated nonperturbatively. Namely,
a complete functional dependence of $\cH[\Delta(\xv)]$ may need to be
taken into account, and an arbitrary form of $\Delta(\xv)$ (arbitrary
set of Fourier harmonics $\Delta_{\qv}$) must be considered.

The findings of LO~\cite{LO} motivate one to focus on a unidirectional
FFLO order characterized by a colinear set of the $\qv_n$. Under this
restriction, FFLO states fall into two, LO and FF universality
classes. The LO (FF) states are characterized and distinguished by
breaking (preserving) translational and preserving (breaking)
time-reversal symmetries. The low-energy properties of such states can
be well captured with a single Q pair (LO) and a single Q wave vector
(FF) approximations, as we describe below. However, the analysis of
energetics of these states far below $h_{c2}$ requires the
aforementioned fully nonlinear analysis.

In principle this can be accomplished through a BdG
treatment. However, its analytical solution (outside 1d) is only
tractable for a single $\Qv$ FF state. This approach predicts a narrow
sliver of FF state near $h_{c2}$, that closes down at
$-1/\kf\as\approx 0.46$, as the strongly coupled regime is
entered~\cite{SR2006,SR2007}.  The most direct interpretation of this
finding is reflected in our $T=0$ phase diagram in
Fig.\ref{fig:phasediagram}\cite{SR2006,SR2007}, and is quite
pessimistic for the observation of any of the FFLO states due to its
narrow range of stability and the trap induced cloud inhomogeneity.

\begin{figure}[bth]
\vspace{5.5cm}
\centering
\setlength{\unitlength}{1mm}
\begin{picture}(40,45)(0,0)
\put(-18,2){\begin{picture}(0,0)(0,0)
\includegraphics{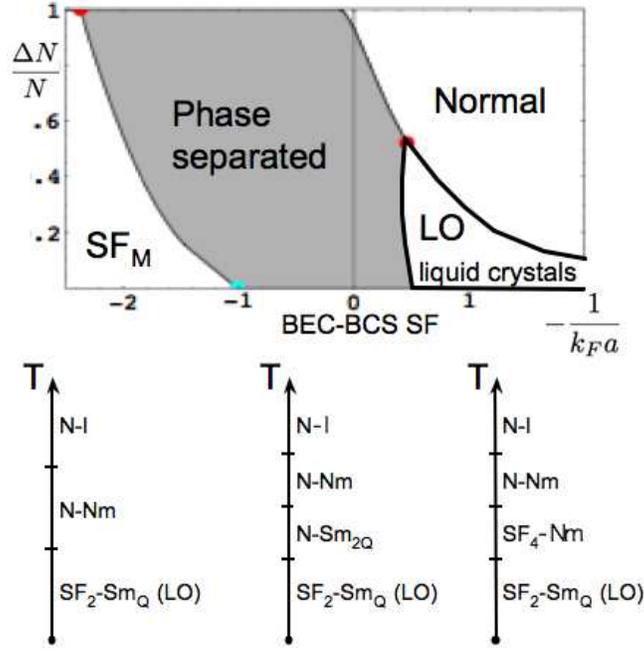}
\end{picture}}
\end{picture}
\vspace{-.5cm}
\caption{Proposed schematic phase diagram of $\Delta N/N$
  vs. $1/(\kf\as)$, showing LO liquid crystal phases replacing
  phase-separated (PS) regime. The three points are the same as in
  Fig. 3, defining the limits of stability of the surrounding
  phases. The lower panel shows symmetry-allowed 3d transition
  scenarios as a function of temperature to the normal-nematic (N-Nm),
  normal-isotropic (N-I), normal-smectic (N-Sm$_{2Q}$), and
  ``charge''-4 superfluid-nematic (SF$_4$-Nm) phases. Currently, no
  microscopic calculations exist that assess which of the three
  possible scenarios are actually realized in imbalanced Fermi gases
  and in what range of detuning.}
\label{phasediagram}
\end{figure}

However, an alternative, more optimistic scenario is suggested by
early numerical treatments of 2d and 3d BdG
equations\cite{MachidaNakanishiLO,Burkhardt94}, exact 1d
solution~\cite{Yang01}, density functional theory
results~\cite{BulgacLO}, and recent findings of a negative domain-wall
energy in an otherwise uniform singlet BCS
superfluid\cite{Matsuo,Yoshida}.  These fully nonlinear (in
$\Delta(\xv)$) findings support the idea that a LO-type periodic
superfluid state of domain-walls (corresponding to a large set of
colinear wavevectors in $\Delta(\xv)$) may be significantly more
stable down to the vicinity of $h_{c1}$.  The LO state can be thought
of as a periodically ordered {\em micro}-phase separation between the
normal and BCS states, that thus naturally replaces the {\em
  macro}-phase separation ubiquitously found in the BEC-BCS
detuning-polarization phase diagram.  In Fig.~\ref{phasediagram}, we
show a proposed phase diagram, assuming that the SF-LO phase
transition is indeed continuous and occupies a wider regime of the
phase diagram than the mean-field prediction (see
Ref.~\cite{Yoshida}).

Despite the extended history of FFLO states, only recently has the
role of quantum and thermal fluctuations been
understood\cite{LeoAshvin}. This study developed Goldstone mode models
for the LO and FF universality classes and analyzed the corresponding
beyond-mean-field phenomenology.  Focusing on the more stable LO
state\cite{LO,MachidaNakanishiLO,Burkhardt94,Yoshida,Matsuo}, it was
demonstrated that in contrast to the conventional (uniform) superfluid
and FF states, the LO superfluid exhibits {\it two\/} Goldstone modes,
$\theta_{\pm Q}$, corresponding to two primary order parameters
$\Delta_{\pm\Qv}$ (with phases for high harmonics locked to these
two). Equivalently, low-energy fluctuations are characterized by the
superfluid phase, $\theta_{sc}=(\theta_++\theta_-)/2$ and the
domain-wall phonon, $u=\theta_{sm}/Q=(\theta_+-\theta_-)/2Q$, with the
LO order parameter (in a single $\pm Q$ approximation) given by
\begin{eqnarray}
\hspace{-1cm}
\Delta_{LO}(\xv)
&=&2\Delta_Qe^{i\theta_{sc}(\xv)}
\cos\big(\qv\cdot\xv + \theta_{sm}(\xv)\big).
\label{LOop}
\end{eqnarray}
The LO order parameter is a product of a superfluid order parameter
and a unidirectional spontaneously oriented (along $\Qv$)
Cooper-pair density wave, i.e., it simultaneously exhibits the
off-diagonal long-range order and smectic order.

Symmetry arguments supported by detailed microscopic
calculations~\cite{LeoAshvin} show that the low-energy modes of the LO
state involve two coupled smectics (with the layer normal taken to lie
along $\xv_\parallel$), with moduli derivable from the BCS theory:
\begin{eqnarray}
  \cH_{LO} &=&\frac{K}{2}(\nabla^2 u)^2 + 
  \frac{B}{2}\bigg(\partial_\parallel u + \frac{1}{2}(\nabla u)^2\bigg)^2
  +\frac{\rho_s^i}{2}(\nabla_i\theta_{sc})^2,
\end{eqnarray}
a form that is familiar from studies of conventional smectic liquid
crystals, with rotational invariance encoded through the vanishing of
the $(\nabla_{\perp} u)^2$ modulus and the specific form of the
nonlinear elastic terms. The LO state is found to be a highly
anisotropic superfluid with the ratio of superfluid stiffnesses given by
\begin{equation}
\rho_s^\perp/\rho_s^\parallel=
\frac{3}{4}\left(\Delta_Q/\Delta_{\rm BCS}\right)^2
\approx\ln(h_{c2}/h)\ll 1,
\label{ratio}
\end{equation}
vanishing for $h\rightarrow h_{c2}^-$.  The FF state is found to be
even more exotic, characterized by an identically vanishing transverse
superfluid stiffness, a reflection of the rotational invariance of the
spontaneous current to an energy-equivalent ground state.
 
Thus, a resonant imbalanced Fermi gas confined to an isotropic trap
is a natural realization of a quantum (superfluid) liquid crystal
that, unlike the solid state analogs, is not plagued by the underlying
lattice potential that explicitly breaks continuous spatial
symmetries.

It has been demonstrated~\cite{LeoAshvin} that in 3d the long-range LO
order is stable to quantum fluctuations, but is marginally unstable at
any nonzero $T$.  The resulting superfluid state is an {\em algebraic}
phase, characterized by universal {\em quasi}-Bragg peaks and
correlations that admit an asymptotically exact description. In
contrast, putative crystalline LO phases\cite{Bowers} with multiple
noncolinear ordering wavevectors are stable against thermal
fluctuations.  The resulting state is also found\cite{LeoAshvin} to
exhibit an unusual topological excitation -- a half vortex bound to a
half dislocation -- allowed by the above form of $\Delta(\xv)$ in
\rfs{LOop}.

In 2d, at nonzero $T$, the state is also an {\em algebraic} phase,
exhibiting universal power-law phonon correlations, controlled by a
nontrivial exactly calculable fixed point. However, it displays
short-range positional order with Lorentzian structure-function peaks,
and is thus unstable to the proliferation of dislocations. The resulting
state is either a ``charge''-$4$ (four-fermion condensate) superfluid
(with order parameter
$\Delta_{sc}^{(4)}\sim\Delta^2\approx\oh\Delta_Q^2e^{i2\theta_{sc}}$),
or a non-superfluid nematic, depending on the relative energetics of
the aforementioned integer and half-integer vortex-dislocation
defects.  The latter normal nematic state is a (complementarily
described~\cite{RadzihovskyDorsey}) deformed Fermi surface
state~\cite{Oganesyan,Sedrakian}, that is another exotic candidate for
a strongly interacting, imbalanced Fermi gas.

The unbinding of various combinations of topological defects
(dislocations and vortices) yields predictions for a number of other
interesting phases\cite{LeoAshvin}. These include two types of
aforementioned orientationally-ordered but statistically homogeneous
(dislocations are unbound) nematic states, one where superfluidity is
completely destroyed (``normal'' nematic, N-Nm) and another in which
4-atom superfluid order accompanies the orientational nematic order
(``superfluid'' nematic, SF$_4$-Nm). Another state is the ``normal''
smectic phase, N-Sm$_{2Q}$, emerging from the LO state by proliferation
of vortices, that destroy the superconducting phase coherence, but
retain the periodic density layering. The three possible
symmetry-allowed scenarios for transitions between these liquid
crystal phases as a function of temperature are illustrated in the
lower panel of Fig.~\ref{phasediagram}.  Currently, no microscopic
calculations exist that assess which of the three possible scenarios
are actually realized in imbalanced Fermi gases and in what range of
detuning.

A complete low-energy description of these LO liquid crystal states
must take the gapless fermionic excitations into account, in addition
to the above bosonic Goldstone modes. As with other analogous
problems~\cite{Oganesyan}, these are expected to lead to the
Landau-like damping of the Goldstone modes $\theta_{sc},u$, and a
finite fermionic quasi-particle lifetime. So far, this difficult
problem has not been addressed.

\subsection{Experimental predictions}

The possibility of the observation of conventional FFLO states has
been discussed at length~\cite{Mizushima,Yang05,SR2006,SR2007}. Bragg
peaks (reflecting {\it spontaneous\/} periodicity) observed in time of
flight~\cite{SR2007} experiments and the spectra of collective
breathing modes\cite{NigelCooper09} are two prominent signatures. It has
been recently emphasized that the FFLO states may be energetically
stabilized in a quasi-1d geometry\cite{Parish1D}.

In the inhomogeneous environment of the trap there are a number of
constraints on the  observation of FFLO states and the fluctuation
phenomena discussed above. Near $h_{c2}$ the LO period
$\lambda_Q=2\pi/Q_0$ is bounded by the coherence length (that near
unitarity can be as short as $\sim R/N^{1/3}$, where $R$ is the
trapped condensate radius and $N$ is the total number of atoms), and
thus $\ll R$ and in this regime the trap can be treated via a local
density approximation (LDA). For $\lambda_Q\ll R$, LDA predicts a weak
pinning of the LO smectic, that can be estimated via finite size
scaling, with trap size $R$ cutting off $\langle
u^2\rangle\sim\eta\log(R/\lambda_Q)$, leading to
$\langle\Delta_{LO}\rangle\sim (\lambda_Q/R)^\eta\ll 1$ that no longer
truly vanishes although it is  still strongly suppressed.  We expect the
predicted strong-fluctuation effects to be experimentally
accessible. We note, for example, that Kosterlitz-Thouless phase
fluctuation physics has been reported in 2d trapped
superfluids\cite{Hadzibabic06}, despite the finite trap size.
However, a more detailed analysis of the trap effects, necessary near
$h_{c1}$ for a quantitative comparison with experiments, remains to be
done.

Other experimental signatures of the LO state include vortex
fractionalization, where the basic superfluid vortex is half the
strength of a vortex in a regular paired condensate. This should be
observable via a doubling of the vortex density in a rotated state.
Also under rotation, the high superfluid anisotropy Eq.~\rf{ratio} is
expected to lead to an imbalance-tunable strongly anisotropic vortex
core and a lattice that is highly stretched along $\Qv$. Bragg peaks in the
time-of-flight images can distinguish the periodic SF$_2$-Sm$_Q$
(superfluid smectic) state from the homogeneous SF$_4$-Nm (superfluid
nematic), which are in turn distinguished from the N-Sm$_{2Q}$ and
N-Nm (normal smectic and nematic) by their superfluid properties,
periodicity, collective modes, quantized vortices, and condensate
peaks. Thermodynamic signatures can identify the corresponding phase
transitions.

\section{Local density approximation}
\label{sec:lda}
One of the principal experimental probes of ultra-cold atomic gases is
the measurement of the local atom density of the trapped cloud.  Thus,
quite generally, detailed theoretical predictions for the phases and
phenomena of cold polarized Fermi gases must account for the effect of
the trapping potential $V_T(r)$.  In recent years, there have been
numerous theoretical investigations of the density profiles of trapped
polarized fermion gases (see, e.g.,
Refs.~\cite{SR2006,Pieri,Kinnunen,YiDuan,DeSilvaPRA,Imambekov,DeSilva06,Gubbels,SR2007,LiuHu2007,HaqueStoof,Jensen,Mizushima2007,Tempere,Baur});
here we shall focus on the predictions of the simplest method for
handling the trapping potential, namely the local density
approximation (LDA).

Much like the WKB approximation, the LDA
corresponds to using expressions for the bulk system, but with an
effective local chemical potential $\mu(r)=\mu-V_T(r)$ in place of
$\mu$.  Systems for which the LDA holds, then, are of particular
interest as they are relevant for the comparison to correlated
condensed matter.  The validity of the LDA relies on the
smoothness of the trap potential, namely that $V_T(r)$ varies slowly
on the scale of the {\it longest\/} physical length $\lambda$ (the
Fermi wavelength, scattering length, effective range, etc.) in the
problem, i.e., $(\lambda/V_T(r)) dV_T(r)/dr\ll 1$. Its accuracy can be
equivalently controlled by a small parameter that is the ratio of the
single particle trap level spacing $\delta E$ to the smallest
characteristic energy $E_c$ of the studied phenomenon (e.g, the
chemical potential, condensation energy, etc.), by requiring $\delta
E/E_c \ll 1$.

\begin{figure}[tbp]
\epsfxsize=9cm 
\centerline{\epsfbox{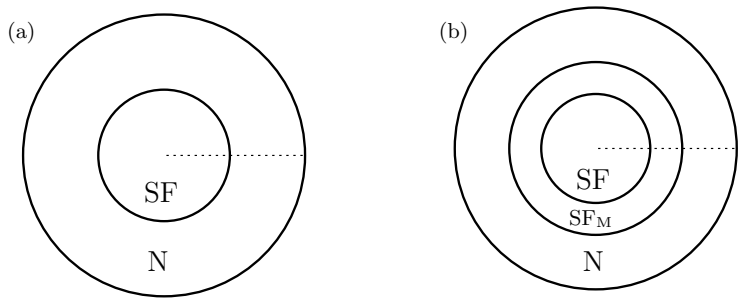}}
\caption{Schematic depiction of the shell structure of polarized
  spherical cold-atom clouds, with the phases occurring with increasing
  radius in a trapped polarized Fermi gas, within the LDA, (a) in the
  BCS and unitary regimes and in (b) the BEC regime where the SF$_{\rm
    M}$ phase exists. }
\label{fig:ldacartoons}
\end{figure}

Because of its more direct current experimental relevance, in this
section we focus on the single-channel model,
Eq.~(\ref{eq:singlechannelintro}).  The generalization of this model
to a trap is straightforward:
\bea
&&\hspace{-1cm} H = \sum_{\sigma=\uparrow,\downarrow}
\int d^3r \ch_\sigma^\dagger(\rv)
\Big[-\frac{\nabla^2}{2m} +
(V_T(\rv) - \mu_\sigma) \Big] \ch_\sigma^\phdag(\rv)
+
\fermiint\int d^3 r\ch_\uparrow^\dagger 
\ch_\downarrow^\dagger \ch_\downarrow^\phdag \ch_\uparrow^\phdag,
\label{eq:singlechannellda}
\eea
where $\ch_\sigma(\rv)$ is a fermionic field operator with Fourier
transform $\ch_{\bk\sigma}$.  Henceforth, to be concrete, we shall
focus on an isotropic harmonic trap $V_T(\rv) = V_T(r)= \frac{1}{2} m
\Omega_T^2 r^2$, although this simplification can be easily relaxed to
handle an arbitrary anisotropic trap. The LDA assumes that, locally,
the system is taken to be well approximated as {\it uniform\/}, but
with a local chemical potential given by \be
\label{eq:mulda}
\mu(r) \equiv \mu - \frac{1}{2} m \Omega_T^2 r^2, 
\ee
where the constant $\mu$ is the true chemical potential (a Lagrange
multiplier) still enforcing the total atom number $N$.  The
spatially-varying spin-up and spin-down local chemical potentials are
then given by
\bea
\mu_\uparrow(r) &=& \mu(r) +h ,
\\
\mu_\downarrow(r) &=& \mu(r) - h,
\eea
with the chemical potential difference $h$ {\it uniform\/}.  Thus,
within the LDA we approximate the energy density by that of a uniform
system with spatial dependence (via the trap) entering only through
$\mu(r)$.  The ground state energy is then simply a volume integral of
this energy density.
Within LDA, the phase behavior as a function of chemical potential,
$\mu$, translates into a spatial cloud profile through $\mu(r)$, with
critical phase boundaries $\mu_c$ corresponding to critical radii
defined by $\mu_c = \mu(r_c,h)$. As we first predicted~\cite{SR2006},
this leads to a shell-like cloud structure (illustrated in
Fig.~\ref{fig:ldacartoons}) that has subsequently been observed
experimentally~\cite{Zwierlein2006,Partridge2006,Shin2006}.

Below we briefly review these shell structures using the LDA,
following our earlier work~\cite{SR2007}. We note, however, that
throughout our discussion, sharp (discontinuous) features that arise
are an artifact of LDA and are expected to be smoothed on microscopic
(Fermi wave-) length scales by the kinetic energy (or, surface
tension~\cite{DeSilva06}).

\begin{figure}[tbp]
\epsfxsize=9cm 
\centerline{\epsfbox{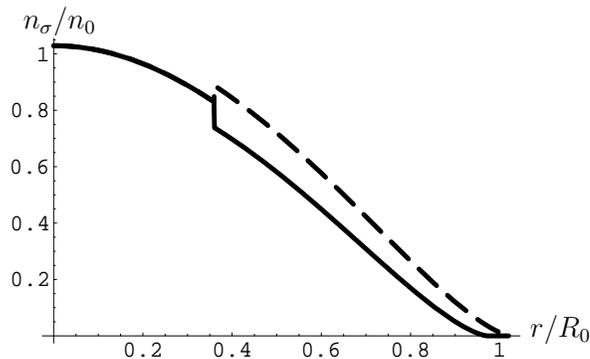}} \vskip-.55cm
\caption{Plot of the local fermion densities, $n_\uparrow$ (dashed) and
  $n_\downarrow$ (solid) , as a function of radius and normalized to the
  overall density $n_0$, as a function of position normalized to the
  $P=0$ Thomas-Fermi radius $R_0$, in the BCS regime with
  $(\kf|\as|)^{-1}=1.5$, from Ref.~\cite{SR2007}. }
\label{fig:nupndownlda}
\end{figure}
We begin with the analysis in the BCS regime, where $\mu>0$.
Qualitatively, the unitary regime will behave analogously to the BCS
regime.  Neglecting the FFLO phase, the fixed $\mu$ and $h$ phase
diagram has a first-order phase transition to the normal phase at
\bea
h_c \simeq \frac{1}{\sqrt{2}} \Delta_{\rm BEC-BCS}(\mu),
\\
\simeq 4\sqrt{2} {\rm e}^{-2} \mu(r) \exp\Big[ \frac{m}{4\pi \as N(\mu(r))}\Big],
\eea
separating a polarized normal phase (when $h>h_c$) and a paired
superfluid phase (when $h<h_c$).  Since $\Delta_{\rm BEC-BCS}(\mu)$ is
an increasing function of $\mu$, this leads to the LDA prediction of a
fully paired superfluid core at small $r$ (where $\mu$ is larger)
surrounded by a population imbalanced normal phase, as illustrated in
Fig.~\ref{fig:ldacartoons}a.  This yields a remarkably simple behavior
for the spin-$\uparrow$ and spin-$\downarrow$ densities
$n_{\uparrow,\downarrow}$, plotted in Fig.~\ref{fig:nupndownlda}, with
$n_\uparrow = n_\downarrow$ in the central core (the fully-paired
superfluid phase), and $n_\uparrow \neq n_\downarrow$ in the outer
shell (the imbalanced unpaired Fermi liquid phase).  We note here that
including the FFLO phase, that occurs near the SF-N phase transition,
would imply a thin shell of FFLO within the LDA yielding oscillations
of the local pair amplitude near the SF-N interface, as also found
in recent beyond-LDA analysis based on the BdG equations~\cite{Kinnunen}.

\begin{figure}[tbp]
\epsfxsize=9cm 
\centerline{\epsfbox{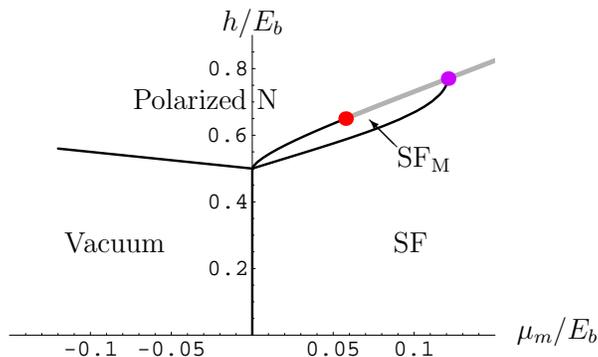}} \vskip-.55cm
\caption{Phase diagram, in the grand-canonical ensemble at fixed $\mu$
  and $h$, with $h$ normalized to the molecular binding energy $E_b =
  \frac{\hbar^2}{m\as^2}$ and $\mu_m = 2\mu +E_b$ the effective
  molecular chemical potential. The thick grey line is a first-order
  transition, and the thin black lines are continuous transitions.
  The red point is a tricritical point and the purple point is a
  critical end point. }
\label{fig:phasebecmuh}
\end{figure}

Turning to the BEC regime, Fig.~\ref{fig:phasebecmuh} shows the
homogeneous-case phase diagram at fixed $\mu$ and $h$ that is required
to construct the phases occurring with increasing radius within the
LDA.  The structure of this phase diagram is somewhat nonintuitive;
this is because at fixed densities the chemical potential in the BEC
regime changes rapidly with external parameters (like the detuning),
as also seen in the balanced case, see Fig.~\ref{fig:becbcs}b.
Nonetheless, the resulting shell structures are the natural
generalization of the BCS regime results: Within the LDA, possible
sequences of phases with increasing radius correspond to constant $h$
trajectories, with decreasing $\mu$, in the phase diagram of
Fig.~\ref{fig:phasebecmuh}. Thus, we see that the possible sequences
of phases with increasing $r$ are: pure SF, SF$\to$ N, or
SF$\to$SF$_{\rm M}$$\to$ N, with the latter case illustrated in
Fig.~\ref{fig:ldacartoons}b.

\begin{figure}[tbp]
\epsfxsize=9cm 
\centerline{\epsfbox{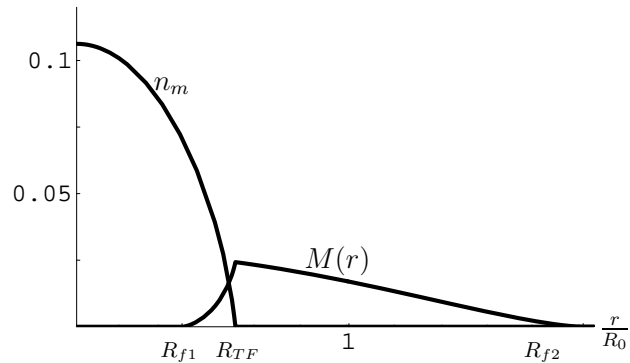}} \vskip1.25cm
\caption{Typical pair density, $n_m \propto |\Delta|^2$, and local
  magnetization, $M$, each normalized to a characteristic density
  scale $E_bN(E_b)$, with $N(E_b)$ the density of states measured at
  the binding energy, from Ref.~\cite{SR2007} in the BEC regime.}
\label{fig:ldabeccurves}
\end{figure}

The characteristic LDA profiles in the BEC regime for the local pair
density $n_m(r)$ and magnetization $M(r)$ are shown in
Fig.~\ref{fig:ldabeccurves} for the sequence of phases SF$\to$SF$_{\rm
  M}$$\to$ N.  Thus, as illustrated in Fig.~\ref{fig:ldacartoons}, in
contrast to the BCS regime here the cloud can exhibit a thin shell of
the \sfm phase with finite polarization and a pairing gap.  The three
characteristic radii of importance are: $R_{f1}$ (the SF$\to$SF$_{\rm
  M}$ transition), $R_{TF}$ (the Thomas-Fermi radius, where the
SF$_{\rm M}$$\to$ N transition occurs) and $R_{f2}$, the outer
boundary of the cloud.
As we discuss in the next section, the LDA has been used by the
Ketterle group~\cite{ShinNature} to infer the phase diagram for a
homogeneous system.  However, interesting physics also holds for
systems in which LDA is violated.  As discussed by DeSilva and
Mueller~\cite{DeSilvaPRA}, the early Rice
experiments~\cite{Partridge2006} showed axial density profiles
(corresponding to the density, integrated over two directions, as a
function of the coordinate $z$ along the long axis of the cloud) that
apparently violated the LDA.  Indeed, the violation of the LDA can be
directly seen in the data shown in Fig.~1 of
Ref.~\cite{Partridge2006}, since the fermion densities do not follow
the contours of the large aspect ratio trap.  Additionally DeSilva and
Mueller showed (see also Ref.~\cite{Imambekov}) that for LDA-type
cloud shapes of the form of Fig.~\ref{fig:ldacartoons}, the axial
magnetization must decrease monotonically with increasing $z$.  A
principal culprit in the breakdown of LDA is the surface tension
associated with the SF-N interface~\cite{DeSilva06}.  This issue was
studied in detail in Ref.~\cite{Baur} in an attempt to understand the
density profiles from the Rice
group~\cite{Partridge2006,Partridge2006prl}. However, it was found
theoretically that although surface tension indeed deforms the
minority cloud in a way qualitatively consistent with the experiments,
the size of the deformation is too small to account for experimental
observations based on microscopic estimates of surface tension
magnitude.  This leaves a quantitative understanding of the Rice
experiments an open question; one possible important ingredient that
needs to be included are fluctuations around mean-field theory (known
to be important in the unitary regime), as discussed in recent
work~\cite{Tempere}.

\section{Recent experiments}
\label{SEC:recentexp}
We now turn to a more comprehensive discussion of recent experimental
developments in imbalanced Feshbach-resonant fermionic atomic gases.
These have primarily been led by the Ketterle group at MIT and the
Hulet group at Rice, launched by two studies of $^6$Li that appeared
in Science in 2006~\cite{Zwierlein2006,Partridge2006}.

The MIT experiments~\cite{Zwierlein2006} reported a destruction of
superfluidity (probed by the appearence of vortices in response to an
imposed rotation) and condensation (probed by a fit to the cloud shape
after sweeping to the BEC side of the resonance) with increasing
imbalance.  As shown in Fig.~1 of Ref.~\cite{Zwierlein2006}, a rotating
polarized Fermi gas exhibits a distinctive inner superfluid core
(defined by a region to which the induced vortex lattice is confined)
surrounded by a vortex-free shell of imbalanced normal cloud.
Consistent with this identification, Fig.~2 of
Ref.~\cite{Zwierlein2006} shows that the number of vortices $N_v$
decreases with increasing imbalance for different values of the
detuning.  On the basis of the standard relation~\cite{Feynman} $N_v =
2m\Omega A/\pi \hbar$, we thus observe that $N_v$ simply measures the
cross-sectional area $A$ (normal to the axis of rotation) of the
superfluid core as a function of the rotation rate $\Omega$. Although
qualitatively this data (the shrinking of the superfluid core with
increasing imbalance) is consistent with current theoretical
predictions, these observations remain to be quantitatively
understood.  Zwierlein, et al.~\cite{Zwierlein2006} also reported
evidence of phase separation, as reflected in a suppression of the
local polarization (species imbalance) near the center of the cloud, a
signature of a core that is fully-paired into a BEC-BCS singlet
superfluid. These were confirmed by in situ measurements of individual
species density profiles using an exquisite phase-contrast imaging and
3d image reconstruction technique\cite{Shin2006}. Based on these
observations they constructed a phase diagram~\cite{Zwierlein2006} with a phase boundary
separating the normal phase and superfluid phases (the latter identified
with normal-superfluid phase separation), in qualitative agreement
with the mean-field theoretical phase diagram~\cite{SR2006,SR2007}.

The initial Rice experiments~\cite{Partridge2006} studied the
imbalanced $^6$Li Fermi gas by comparing the density profiles of the
two fermion species. Consistent with the MIT experiments they found
that, with increasing species number imbalance, the imposed imbalance
is pushed to the outside shell of the cloud.  Furthermore, by
measuring the dependence of the radii of the Thomas-Fermi clouds for
the two fermion species and showing that the ratio of the majority to
minority radius, $R_{\uparrow}/R_{\downarrow}$, increases with
increasing imbalance $P$, Partridge, et al. showed evidence for phase
separation in the unitary regime. From these measurements, the Rice
group found a putative {\em nonzero} lower-critical polarization for
phase separation with $P_{c1}\simeq 0.09$, observing that for
$P<P_{c1}$ the radii of the two species coincided,
$R_{\uparrow}=R_{\downarrow}$, despite a nonzero imposed imbalance
(see Fig. 3 of Ref.~\cite{Partridge2006}).  We are not aware of a
ground state (within a comprehensive theoretical picture) that can
accomodate this observation of a nonzero $P_{c1}$, i.e., a state that
is both fully paired and imbalanced; naive candidates like the homogeneous
polarized superfluid SF$_M$ (appearing on the BEC
side)~\cite{SR2006,SR2007} and a putative $p$-wave paired
state~\cite{BulgacPwavePRL} would fail toward the edge of the cloud,
where on general grounds a polarized normal state with distinct radii
is expected. Furthermore, a reasonable understanding of this data was
obtained by Chevy~\cite{Chevy} within a simple unitary model of phase
separation (i.e., with distinct majority-minority radii) between a
fully-paired unpolarized superfluid and a fully-polarized normal
state, using universal forms for the equations of state for these
phases. Taking these together with the fact that no such finite
$P_{c1}$ was ever observed by the MIT group~\cite{Shin2006} suggests
that the putative (small) nonzero $P_{c1}$ in Rice measurements may be
an experimental artifact. An alternative possibility is that it is a
consequence of a nonzero temperature $T\approx 0.1 T_F$, that on one
hand allows a finite polarization (through thermally activated
quasi-particles) and on the other entropically suppresses phase
separation, thereby allowing a nonzero $P_{c1}$. One support for the
latter scenario is the fact that in the later Rice
experiments~\cite{Partridge2006prl} a vanishing $P_{c1}$ was found at
reduced temperature, $T\approx 0.05 T_F$. Detailed experiments further
clarifying this issue would be highly desirable.

A number of subsequent experiments further explored the onset of phase
separation in such imbalanced Fermi gases.  Zwierlein, et
al.~\cite{Zwierlein2006} studied the transition between the unpaired
and superfluid phases by observing column density profiles in
imbalanced Fermi gases, finding the upper-critical polarization of
$P_{c2,{\rm trap}} \simeq 0.70$ at the unitary point for the {\it
  trapped\/} cloud.  Due to the presence of the trap, this value
cannot be directly compared to the upper-critical polarization
$P_{c2}$ for the bulk, uniform system, the latter estimated within
mean-field analysis to be $P_{c2}^{\rm mft}\simeq 0.93$.  However,
within the LDA, quite generally
\be
 P_{c2,{\rm trap}}>P_{c2}.
\label{eq:trappol}
\ee
This bound that can be understood by considering a trapped imbalanced
Fermi gas at $P_{c2,{\rm trap}}=M/n = \Delta N/N$, with magnetization
$M=\Delta N/V$ ($V$ the system's volume) and atom density $n = N/V$.
Within the LDA, at $P_{c2,{\rm trap}}$, by definition the fully-paired
inner superfluid core has shrunk to zero radius. Thus, at the center
of the trap the local polarization is given by the bulk upper-critical
polarization, i.e., $P(r= 0) =M(0)/n(0) = P_{c2}$, where $M(r)$ is the
local magnetization and $n(r)$ is the local density.  Now, since
$\mu(r) = \mu - \frac{1}{2}m\Omega^2 r^2$ decreases with increasing
radius, and the chemical potential difference $h$ is $r$-independent,
$M(r)$ and $n(r)$ respectively increase and decrease with increasing
radius, implying
\bea
\Delta N &=& \int d^3 r M(r) \geq V M(0),
\\
N &=& \int d^3 r n(r) < Vn(0).
\eea
and giving the bound in \rfs{eq:trappol}.  Thus, the observation of
$P_{c2,{\rm trap}} \simeq 0.70$ in Ref.~\cite{Zwierlein2006} is a
measure of a lower-bound on the inaccuracy of the mean-field estimate
for $P_{c2}$.

The imaging technique in such experiments naturally integrates the
local atomic density along the axis perpendicular to the camera's
imaging plane. To deconvolve this projection, Shin, et
al.~\cite{Shin2006} used the inverse Abel
transform~\cite{KetterleZwierlein} to extract the local
three-dimensional atom densities.  This development allowed a clear
identification of the paired superfluid core (within which the local
imbalance identically vanishes, at $T=0$) in the phase separation
regime, and the measure of the upper-critical polarization as a
function of detuning, finding $P_{c2,{\rm trap}} = 0.77$ at unitarity,
close to unity in the BEC regime, and decreasing in the BCS regime.

One of the most significant differences between the MIT and Rice
observations is the shape of the interface between the balanced
superfluid core and the imbalanced outside shell. While the MIT group
found an interface whose shape mimics the trap (see, e.g., Fig.~6 of
Ref.~\cite{Shin2006}), the Rice group observed that the imposed
imbalance is accomodated by an enhanced local polarization in the
axial poles of the cloud~\cite{Partridge2006}, thereby at low $T$
deviating from the trap's highly anisotropic shape.  This discrepancy,
which to date remains to be quantitatively understood, is believed to
be related to the very different trap aspect ratios in the MIT
($f_r/f_z = 5.6$ in Ref.~\cite{Shin2006}) and Rice ($f_r/f_z = 49$ in
Ref.~\cite{Partridge2006}) experiments, the latter exhibiting a nearly
one-dimensional trap geometry.  This issue was studied in more detail
in Ref.~\cite{Partridge2006prl}, where the distorted shape
(attributable to a nonzero surface tension~\cite{DeSilvaPRA}
associated with the first-order transition and corresponding phase
separation) of the paired inner core was confirmed, and the dependence
of the aspect ratios of the minority and majority species fermionic
clouds as a function of the imposed imbalance was measured.  These
later Rice experiments~\cite{Partridge2006prl} also found a
surprisingly large value of the in-trap upper critical polarization,
$P_{c2,{\rm trap}}\approx 0.9$.  Subsequent Rice
experiments~\cite{Hulet_unpublished} explored the effects of trap
aspect ratio, and found that the LDA-violating deformations decrease
with decreasing aspect ratio, with no deformations observed for aspect
ratios below $6$.  Furthermore, the critical polarization was found to
be approximately $P \approx 0.75$, consistent with the MIT measurements and
Monte Carlo calculations.

However, recent Paris experiments~\cite{Nascimbene} studied the phase
diagram of an imbalanced Fermi gas with around $10^5$ atoms and trap
aspect ratio of around $20$, conditions in between the Rice and MIT
systems. In contrast to the above appealing resolution of the discrepancy between the 
Rice and MIT observations (due to a very different aspect ratio),
these Paris experiments saw no appearance of LDA-violating features or
surface tension effects. Furthermore their results are in a very good
agreement with MIT's, both qualitatively (observing a flat-top
distribution of the density difference, that can arise from a fully-paired
superfluid core within the LDA), and quantitatively (with respect to the critical
polarization $P_{c2}$). Thus, a resolution of these puzzling
discrepancies remains an open issue.

The observed matching of the majority and minority clouds' aspect
ratios to the trap's aspect ratio in the MIT experiments indicates
that, for these experiments, the LDA is well satisfied.  The MIT group
then took advantage of the LDA to extract the $T=0$ and finite $T$
phase diagrams~\cite{ShinNature,ShinPRL2008} for a bulk system,
facilitating a more direct comparison with theoretical studies of a
homogeneous gas.  Focusing on the unitary limit, these experiments
found a finite-$T$ phase diagram of the standard tricritical form (i.e.,
like Fig.~\ref{fig:phasediagramT}, resembling other tricritical
systems, such as $^3$He-$^4$He mixtures~\cite{Graf67},
metamagnets~\cite{Shang80}, and thin-film
superconductors~\cite{WuAdams}).

As illustrated in Fig.~\ref{fig:phasediagramT}, a tricritical point
separates a second-order (at high $T$) from a first-order (at low $T$)
transition that, in the present case, is between a paired superfluid
and a polarized normal state, with the latter extending to a regime of
phase separation at fixed polarization.  This general
structure is well captured by mean-field theory for the
narrow-resonance model~\cite{Parish2007}, however the shape of the
phase diagram in the experimentally-relevant wide-resonance regime
remains an open question.  Finite temperature analysis beyond
mean-field theory~\cite{Chien2007} finds a more complicated phase
diagram topology, with the regime of phase separation persisting for
temperatures higher than the tricritical point.

The principal experimental evidence for the expected tricritical point
comes from the observations~\cite{Partridge2006prl,ShinNature} that at
higher temperatures the breakdown of LDA (as signaled by the distinct
cloud and trap shape anisotropies) ceases. This was reasonably
interpreted as the disappearance of the surface tension (associated
with phase coexistence at first-order transition) as the low-$T$
first-order superfluid-normal transition converts to a continuous one
for $T$ above the tricritical temperature. This is further supported
by observations that at these higher temperatures (above the
tricritical temperature) the expected Clogston limit of $P_{c2,{\rm
    trap}}\approx 0.75$ is recovered. A more detailed check on the
consistency of the interpretation can come from a comparison with the
predictions for the cloud shape, heat capacity, and local polarization
controlled by the mean-field tricritical point~\cite{SheehyTri}, which
are predicted to follow distinct power laws~\cite{Stephen75,Lawrie}.

As mentioned above, in later work~\cite{ShinPRL2008}, Shin, et
al. also used the LDA to extract the $T=0$ phase diagram of an
imbalanced Fermi gas. The resulting experimental phase diagram agrees
qualitatively with the theoretical one displayed in
Fig.~\ref{fig:phasediagram}~\cite{SR2006,SR2007}, though no
experimental evidence for the putative FFLO phase exists to date, nor
has the low-$T$ SF$_{\rm M}$-to-PS phase boundary ($P_{{\rm c}1}$)
been mapped out.  The observed value of $P_{c2}\simeq 0.36$ at
unitarity (reported in Ref.~\cite{ShinNature}, with the phase diagram
of Ref.~\cite{ShinPRL2008} showing a slightly lower value) is
consistent with the above bound $P_{c2}<P_{c2,{\rm trap}}$ and points
to a large quantitative error in the mean-field estimate (although
including the leading $1/N_f$ correction~\cite{Veillette07} yields a
more accurate estimate, $P_{{\rm c}2} \simeq 0.302$) This measurement
is consistent with the quantum Monte Carlo based prediction by Lobo,
et al~\cite{lobo2006}, who find $n_\uparrow/n_\downarrow = 0.44$ at
the transition, corresponding to $P_{c2}= 0.38$ and within LDA giving
a universal value of $P_{c2,{\rm trap}} = 0.77$, in agreement with the
MIT~\cite{Shin2006} and latest Rice~\cite{Hulet_unpublished}
measurements.

One remarkable aspect of the experiments in Ref.~\cite{ShinPRL2008} is
that the density profiles of the trapped imbalanced Fermi gases were
quantitatively captured within a simple boson-fermion mixture model
(with bosons representing the tightly bound molecular pairs and
fermions representing the excess spin-up fermions), using the vacuum
molecule-fermion~\cite{Skorniakov} ($a_{bf} = 1.18a$) and
molecule-molecule~\cite{Petrov2004,Levinsen2005} ($a_{bb} = 0.6 a$)
scattering lengths.  Since the latter values are only valid in the
dilute deep-BEC regime, it is surprising that such an accurate fit to
the data held close to unitarity.  It will be interesting, in future
work, to test the limits of the boson-fermion model used in
Ref.~\cite{ShinPRL2008} and to look for deviations from this simple
picture.

Before turning to the last topic of this section, which is the
development of RF spectroscopy probes of imbalanced Feshbach-resonant
Fermi gases and the strongly-imbalanced polaron regime, we briefly
comment on recent Rice experiments~\cite{LiaoRittner}, that have
attained the quasi one-dimensional conditions for which the 1d FFLO
state is expected.  These experiments utilize a two-dimensional
optical lattice to create an array of weakly-coupled, large aspect
ratio tubes (with $\omega_r/\omega_z = 1000$), that are therefore
effectively in a one-dimensional limit.  Previous theoretical
work~\cite{Parish1D} and general arguments suggest that in this limit
the FFLO state should robustly appear. Moreover, uniform
one-dimensional imbalanced Fermi gases can be solved exactly via Bethe
ansatz~\cite{Orso,HuLiu}, supporting these claims.  Liao et
al~\cite{LiaoRittner} found that, combining such Bethe ansatz results
with LDA (to account for the spatial dependence along the tube axis),
a quantitative agreement with their trapped density profiles was
obtained.  Consistent with earlier theoretical
predictions~\cite{Parish1D}, these experiments found that in contrast
to the 3d case, the superfluidity is weakest and the imbalance
(magnetization) is largest at the {\it center\/} of the trap where the
chemical potential is smaller. This is a consequence of the reduction
in the 1d density of states with the increasing chemical potential
(contrasting with the opposite behavior in 3d), resulting in the
appearance of the balanced superfluid at small $\mu$, and the
imbalanced superfluid (the 1d equivalent of the FFLO state) at large
$\mu$ at the center of the trap.  An important goal for future work is
a direct observation of the expected modulation of the FFLO
state. This should be reflected in quantities such as the local
pairing correlations, as found in recent QMC~\cite{Batrouni07} and
density-matrix renormalization group~\cite{Feiguin,Tezuka} studies of
one-dimensional systems. Finite momentum Bragg BEC peaks (akin to
those appearing in a condensates in an optical lattice, but here
without an imposed periodic potential) are also expected to
characterize the time-of-flight measurements of the FFLO states.

We now discuss RF spectroscopy~\cite{Chin2004}, which has been
fruitfully used to probe resonant imbalanced Fermi gases, with the
particular focus on the strongly interacting normal state driven by a
high imbalance and/or increased
temperature~\cite{Schunck2007,Schunck2008}.  RF spectroscopy studies
of strongly imbalanced Fermi gases have a close condensed-matter
counterpart, namely tunneling in the paramagnetic phase above the
Clogston limit in thin-film superconductors~\cite{Aleiner,Xiong}.
Instead of removing or adding electrons through an insulating barrier,
in RF spectroscopy the RF field induces transitions from one of the
two interacting hyperfine species of the atomic cloud to a third
unpopulated hyperfine state, and this rate as a function of the RF
frequency $\omega$ is measured by detecting the number $N_3(\omega)$
of the third species. In the simplest interpretation the signal
$N_3(\omega)$ measures the spectral function of the species undergoing
a transition to the third hyperfine state.

However, as was recently
emphasized~\cite{Veillette08,YuBaym,PunkRF,BaymPethick,PeraliRF,Basu,HeRF},
this interpretation is complicated by final state interactions (a
crucial issue that has received much recent theoretical
attention~\cite{Veillette08,YuBaym,PunkRF,BaymPethick,PeraliRF,Basu,HeRF})
of the third hyperfine state that are strong in
$^6$Li~\cite{Schunck2007,Schunck2008}, and a significantly more
involved theoretical analysis is required. (In contrast, in $^{40}$K
final state interactions are weak and a direct comparison with the
fermionic spectral function is possible~\cite{StewartGaeblerJin}).

The key features of the RF signal are the width and the shift of the
peak in $N_3(\omega)$ relative to that in a cloud of noninteracting
atoms. Based on earlier studies\cite{Chin2004} the observed peak shift
was originally associated with the pairing gap. An observation of a
such shift for a strongly imbalanced gas in the regime ($T$,$P$,$k_F
a$) of the phase diagram, where the gas did not display a condensate
peak, was taken as evidence for an exotic paired non-superfluid
state\cite{Schunck2007}.  This is reminiscent of photoemission in the cuprate
high-temperature superconductors, which exhibit a
pseudogap~\cite{Timusk}, loosely taken as evidence of local pairing
correlations~\cite{Emery,Balents,FranzTesanovic,Chen2005} in the
non-superconducting state above $T_c$.  However, a detailed
analysis\cite{Veillette08} (including final state interactions and
controlled by a systematic
$1/N_f$-expansion\cite{Nikolic2006,Veillette07}) of the imbalanced
resonant gas demonstrated that the observed RF shift and widths can be
qualitatively understood in terms of a conventional Fermi-liquid
state, albeit a strongly renormalized one. The latter is characterized
by large self-energy corrections leading to a modified Migdal
discontinuity in the momentum distribution function (single-particle
residue), $Z<1$, the enhanced effective mass, $m^* > m$, and a large
chemical potential shift responsible for the observed shift in the RF
peak\cite{Veillette08}. Subsequent experiments from the same
group\cite{Schunck2008} have successfully reinterpreted the original
observations in terms of this theoretically advocated Fermi-liquid
picture, though a detailed quantitative description of the data is
still lacking.

Finally, as mentioned earlier recent experimental
studies~\cite{Schirotzek2009,Nascimbene} have provided additional
insight on this interesting system in the strongly imbalanced limit of
$P\to 1$, corresponding to a single minority (e.g., spin down) atom
moving and resonantly interacting with a fully polarized Fermi sea of
majority atoms (spin up).  Schirotzek, et al~\cite{Schirotzek2009}
find evidence for the predicted
transition~\cite{ProkofevPolaron,Punk2009} from a polaron-like state
of the minority atom interacting with the majority Fermi sea,
characterized by a nonzero fermionic pole with a residue $Z<1$, to a
diatomic bosonic molecule in the Fermi sea, with $Z\to 0$. Subsequent
work by Nascimbene, et al.~\cite{Nascimbene} studied the collective
oscillations of the strongly imbalanced Fermi gas, extracting the
effective polaron mass $m^* \simeq 1.17m$. These measurements are in
good agreement with simple variational calculation based on the Chevy
ansatz~\cite{Chevy,combescot2007} and diagrammatic
Monte-Carlo\cite{ProkofevPolaron}, that both give $m^*=1.17m$, and another
recent calculation~\cite{combescot2007} that yields $m^*=1.20m$.

\section{Concluding remarks and future outlook}
 \label{SEC:concluding}
In this brief review, we have discussed recent developments in the
study of imbalanced resonant Fermi gases. As we described, their
phenomenology is summarized by the phase diagram as a function of
imbalance (polarization $P$) and interaction ($1/k_F a$, controlled by
FR detuning). The zero-polarization paired-superfluid ground state,
exhibiting the well-studied BEC-BCS crossover, is readily destablized
by an imposition of a finite species imbalance, on the BCS side, to a
regime of phase separation between the paired superfluid and a
polarized normal state. In contrast, sufficiently deep in the BEC
regime side this is preempted by a polarized homogeneous superfluid,
consisting of molecular Bose-condensed pairs and a fully polarized
Fermi sea of the remaining majority-species atoms. Beyond the
so-called Clogston limit of the upper-critical imbalance, the strongly
interacting normal state is obtained. The properties of these phases
and regimes have been vigorously studied through measurements of the
number and imbalance density profiles, collective modes, as well as the
RF spectroscopy as a function of interaction strength (controlled by
FR detuning), imbalance and temperature, finding qualitative and at
times quantitative agreement with theoretical predictions.

Despite this coherent picture that has emerged, there are number of
interesting issues that remain to be explored. Firstly, more
detailed and quantitative studies of the phase boundaries are
needed, in particular a determination of the SF$_M$ phase boundary
$P_{c1}$ and the upper and lower critical polarizations as a
function of detuning and temperature.  In particular, it is not known
whether the finite-$T$ phase diagram has the standard tricritical form
for all detunings.

Secondly, more detailed investigations of phases are needed. As
discussed in the manuscript, the polarized superfluid SF$_M$,
appearing in the BEC regime at $T=0$, is described, at the simplest level, by
a simple Bose-Fermi mixture of molecular pairs and
single-species Fermi sea. It would be interesting to explore further
the extent to which this picture is only an approximation and what the
limit of its validity is, particularly as the system is taken closer
to the unitary point, where $s$-wave scattering length grows and the
molecules can no longer be treated as point bosons.

There is also much that remains to be elucidated about the nature of
the strongly interacting normal state at high
polarization. Undoubtedly, RF spectroscopy and its recently-developed
momentum-resolved extension~\cite{StewartGaeblerJin} will be central
to detailing its properties as a function of detuning and temperature.

But probably the biggest remaining enigma is the long sought-after
FFLO state, or more generally an imbalance-driven paired superfluid
state that spontaneously breaks translational and orientational
symmetries. On the theoretical side, the form of the periodic state
(unidirectional or crystalline and of what type~\cite{Bowers}), the
range of stability of the state in the
detuning-polarization-temperature phase diagram, as well as the nature
of the transition to it from the paired BEC-BCS superfluid remain
as important open questions. Furthermore, as recently demonstrated the
simplest LO state admits fractional topological defects (a 1/2-vortex
bound to a 1/2-dislocation), and exhibits enhanced fluctuations that
lead to a variety of other quantum liquid crystal phases and other
rich phenomena\cite{LeoAshvin}.

As discussed above, reliable analytical treatments around
$h_{c2}$\cite{LO} and those specializing to the FF (single $Q$)
state\cite{FF}, and more recent ones that extend these beyond the deep
BCS regime\cite{SR2006,SR2007} suggest that the FFLO-type states are
confined to a narrow sliver near $h_{c2}$, just below the normal
state. Whether this is generic or an artifact of current
approximations remains an open question.

One of the reliable limits that sheds some light on this question is
the case of one dimension, where there is an exact solution of the
Bogoliubov-deGennes equations, predicting a LO type ground state over
a large portion of the phase
diagram\cite{MachidaNakanishiLO,Orso,HuLiu}. 
This one-dimensional regime is also particularly amenable to numerical treatments~\cite{Batrouni07,Feiguin,Tezuka,Batrouni2,Casula}
and exact Bethe-ansatz methods~\cite{Orso,HuLiu}.
 In fact in one dimension, on
general grounds the LO type state is generic for a finite imbalance,
indistinguishable from the 1d SF$_M$ ground state. Thus, the existence
of a FFLO type state in one dimension is theoretically indisputable. Stimulated
by this and supported by quasi-1d mean-field analysis\cite{Parish1D},
recent efforts have focused on a realization of the FFLO state in an
array of weakly coupled 1d imbalanced resonant Fermi gases, created by
a strong 2d optical standing wave
potential\cite{Hulet_unpublished}. In the time of flight, this state
is expected to exhibit spontaneous Bragg peaks associated with the
simultaneous presence of superfluid coherence and periodic positional
order. Undoubtedly, many unanticipated surprises will emerge as a
result of these and many other future studies.

{\em Acknowledgements\/} We gratefully acknowledge past collaborations
with V. Gurarie, A. Lamacraft, E.G. Moon, S. Sachdev, and
M. Veillette, on which parts of this review are based, and stimulating
discussions with D. Jin, E. Cornell, R. Hulet, and  N. Navon, as well as financial
support from NSF DMR-0321848 and the Louisiana Board of Regents, under
grant No.  LEQSF (2008-11)-RD-A-10.



\begin{thebibliography}{10}
\bibitem{Polnote}
In the literature a variety of notations are used to describe the 
imposed population imbalance, including  
the most
physical hyperfine species population imbalance (or simply imbalance),
polarization and magnetization. We will use these terms  
interchangeably.
\end{thebibliography}
\end{document}